\documentclass[fleqn,usenatbib]{mnras}

\usepackage{newtxtext,newtxmath}

\usepackage[T1]{fontenc}

\DeclareRobustCommand{\VAN}[3]{#2}
\let\VANthebibliography\thebibliography
\def\thebibliography{\DeclareRobustCommand{\VAN}[3]{##3}\VANthebibliography}

\usepackage{graphicx}	
\usepackage{amsmath}	
\usepackage{xcolor}
\usepackage{pdflscape}
\usepackage{orcidlink}
\usepackage{comment}
\usepackage{multirow}

\newcommand{\Ha}{H$\alpha$}

\newcommand{\bursteta}{$\eta_{1500}$}

\title[
Burstiness and interactions in low-mass galaxies]{Bursty star formation and galaxy-galaxy interactions in low-mass galaxies 1 Gyr after the Big Bang}

\author[Asada et al.]{
Yoshihisa Asada$^{1,2}$\thanks{e-mail: asada@kusastro.kyoto-u.ac.jp}\orcidlink{0000-0003-3983-5438},
Marcin Sawicki$^{1}$\thanks{Canada Research Chair}\orcidlink{0000-0002-7712-7857},
Roberto Abraham$^{3,4}$\orcidlink{0000-0002-4542-921X},
Maru\v{s}a Brada\v{c}$^{5}$\orcidlink{0000-0001-5984-0395},
Gabriel Brammer$^{6,7}$\orcidlink{0000-0003-2680-005X},\newauthor
Guillaume Desprez$^{1}$\orcidlink{0000-0001-8325-1742},
Vince Estrada-Carpenter$^{1}$\orcidlink{0000-0001-8489-2349},
Kartheik Iyer$^{8}$\orcidlink{0000-0001-9298-3523},
Nicholas Martis$^{1,9}$\orcidlink{0000-0003-3243-9969},\newauthor
Jasleen Matharu$^{6,7}$\orcidlink{0000-0002-7547-3385},
Lamiya Mowla$^{10}$\orcidlink{0000-0002-8530-9765},
Adam Muzzin$^{11}$,
Ga\"el Noirot$^{1}$,
Ghassan T. E. Sarrouh$^{11}$\orcidlink{0000-0001-8830-2166},\newauthor
Victoria Strait$^{6,7}$\orcidlink{0000-0002-6338-7295},
Chris J. Willott$^{9}$\orcidlink{0000-0002-4201-7367},
Anishya Harshan$^{5}$\orcidlink{0000-0001-9414-6382}
\\
$^{1}$ Department of Astronomy \& Physics and Institute for Computational Astrophysics, Saint Mary's University, 923 Robie Street, Halifax, NS B3H 3C3, Canada\\
$^{2}$ Department of Astronomy, Kyoto University, Sakyo-ku, Kyoto 606-8502, Japan\\
$^{3}$ David A. Dunlap Department of Astronomy and Astrophysics, University of Toronto, 50 St. George Street, Toronto, Ontario, M5S 3H4, Canada\\
$^{4}$ Dunlap Institute for Astronomy and Astrophysics, 50 St. George Street, Toronto, Ontario, M5S 3H4, Canada\\
$^{5}$ Department of Mathematics and Physics, Jadranska ulica 19, SI-1000 Ljubljana, Slovenia\\
$^{6}$ Niels Bohr Institute, University of Copenhagen, Jagtvej 128, DK-2200 Copenhagen N, Denmark\\
$^{7}$ Cosmic Dawn Center (DAWN), Denmark\\
$^{8}$ Columbia Astrophysics Laboratory, Columbia University, 550 West 120th Street, New York, NY 10027, USA\\
$^{9}$ National Research Council of Canada, Herzberg Astronomy \& Astrophysics Research Centre, 5071 West Saanich Road, Victoria, BC, V9E 2E7, Canada\\
$^{10}$ Whitin Observatory, Department of Physics and Astronomy, Wellesley College, 106 Central Street, Wellesley, MA 02481, USA\\
$^{11}$ Department of Physics and Astronomy, York University, 4700 Keele St. Toronto, Ontario, M3J 1P3, Canada
}

\date{Accepted XXX. Received YYY; in original form ZZZ}

\pubyear{2023}

\begin{document}
\label{firstpage}
\pagerange{\pageref{firstpage}--\pageref{lastpage}}
\maketitle

\begin{abstract}

We use CANUCS JWST/NIRCam imaging of galaxies behind the gravitationally-lensing cluster MACS J0417.5-1154 to investigate star formation burstiness in low-mass ($M_\star\sim10^8\ M_\odot$) galaxies at $z\sim4.7-6.5$.  Our sample of 123 galaxies is selected using the Lyman break selection and photometric emission-line excess methods.
Sixty percent of the 123 galaxies in this sample have H$\alpha$-to-UV flux ratios that deviate significantly from the range of H$\alpha$-to-UV ratio values consistent with smooth and steady star formation histories. This large fraction indicates that the majority of low-mass galaxies is experiencing bursty star formation histories at high redshift. We also searched for interacting galaxies in our sample and found that they are remarkably common ($\sim40\ \%$ of the sample). Compared to non-interacting galaxies, interacting galaxies are more likely to have very low H$\alpha$-to-UV ratios, suggesting that galaxy-galaxy interactions enhance star formation burstiness and enable faster quenching (with timescales of $\lesssim100$ Myr) that follows the rapid rise of star formation activity. Given the high frequency of galaxy-galaxy interactions and the rapid SFR fluctuations they appear to cause, we conclude that galaxy-galaxy interactions could be a leading cause of bursty star formation in low-mass, high-$z$ galaxies. They could thus play a significant role in the evolution of the galaxy population at early cosmological times.
\end{abstract}




\section{Introduction}\label{sec:intro}
Low-mass galaxies ($\lesssim10^{8}M_\odot$) in the high-$z$ Universe are thought to be the main contributor to the star formation rate density and Cosmic ionizing photon budget.  They are viewed as key to reionizing the Universe and keeping it ionized at $z\gtrsim4$ \citep[e.g.,][]{Bouwens2015,Robertson2015,Livermore2017,Ocvirk2021,Iwata2022}.
Thus revealing how these galaxies evolve is one of the key science goals in modern extragalactic astronomy.
Since galaxies evolve  by converting their gas into stars ("star formation activities"), characterizing  their star formation histories (SFHs) sheds light on the assembly of high-$z$ low-mass galaxies in particular, and the cosmological evolution of galaxies in general.

Studies of low-mass galaxies at lower redshifts showed that low-mass galaxies typically have fluctuating, bursty SFHs rather than steady and smooth SFHs \citep[e.g.,][]{Sullivan2001,Boselli2009,Weisz2012,Dominguez2015,Guo2016,Emami2019}.
"Bursty" SFH here refers to episodic bursts of star formation, with each burst characterized by a rapid rise of the star formation rate (SFR) and subsequent fast quenching, both with short time-scales of $<100$ Myrs.
Low-mass galaxies in the high-$z$ Universe in general have also been expected to experience such bursts of star formation, characterized by rapidly rising and declining SFRs, and to mainly evolve through the bursty SFHs.
Indeed, hydrodynamical simulations of high-$z$ dwarf galaxies showed large variations of the SFRs and infer the presence of bursty SFHs \citep[e.g.,][]{Ma2018,Legrand2022,Pallottini2022,Pallottini2023}.
However, it has been impossible to observationally confirm this scenario before the launch of James Webb Space Telescope \citep[JWST;][]{Gardner2023} due to the lack of sensitivity in rest-frame optical observations at these redshifts.

JWST has introduced a new era and enabled us to investigate the star formation activities of high-$z$ low-mass galaxies in detail.
Early results from JWST observations have already identified a number of low-mass galaxies at $z\gtrsim4$ with extremely high equivalent width emission lines \citep[e.g.,][]{Boyett2022,Matthee2022,Asada2023,Jones2023,Stiavelli2023,Williams2023,Withers2023}, which are expected to be seen in galaxies with rapidly rising SFRs \citep[e.g.,][]{inoue_rest-frame_2011}.
Additionally, some JWST spectroscopic observations reported individual examples of low-mass galaxies showing signatures of on-going rapid quenching, weak emission lines and/or clear Balmer break, even at $z\gtrsim4$ \citep[e.g.,][]{Looser2023,Strait2023}.
These results suggest that low-mass galaxies are indeed experiencing  bursty SFHs at $z\gtrsim4$, and bursty star formation can be a key period in the evolution of low-mass galaxies \citep[see also][]{Endsley2023,Looser2023b}.

However, the physical origin of such bursty SFHs in high-$z$ low-mass galaxies is still unclear.
Some hydrodynamical simulations suggest that  clustered star formation and strong stellar feedback trigger the rapid rise and decline of the SFR, respectively, resulting in bursty SFHs.  \citep[e.g.,][]{Hopkins2014,Sparre2017}.
Particularly, the presence of quiescent low-mass galaxies at high-$z$ reported by JWST observation has taken us by surprise, and several simulations discussed the physical origins of the rapid quenching in these high-$z$ low-mass galaxies, suggesting the presence of other mechanisms such as radiation-driven winds \citep{Gelli2023} or galaxy-galaxy interactions \citep{Dome2023}.
Galaxy-galaxy interactions have been proposed as a possible cause of bursty SFHs in the high-$z$ universe in some observational studies as well \citep[][]{Asada2023,Witten2023}, and one of the recently quenched low-mass galaxies at $z\gtrsim4$ indeed has a physically close companion at the same redshift \citep[][and this is the lowest-mass galaxy among the recently-quenched ones reported so far]{Strait2023}.
However, as of yet, a statistical investigation of the correlation between SFH burstiness and galaxy local environments is still lacking.

\subsection{Measuring SFR burstiness using H$\alpha$ and UV}

\begin{figure}
	\includegraphics[width=\columnwidth]{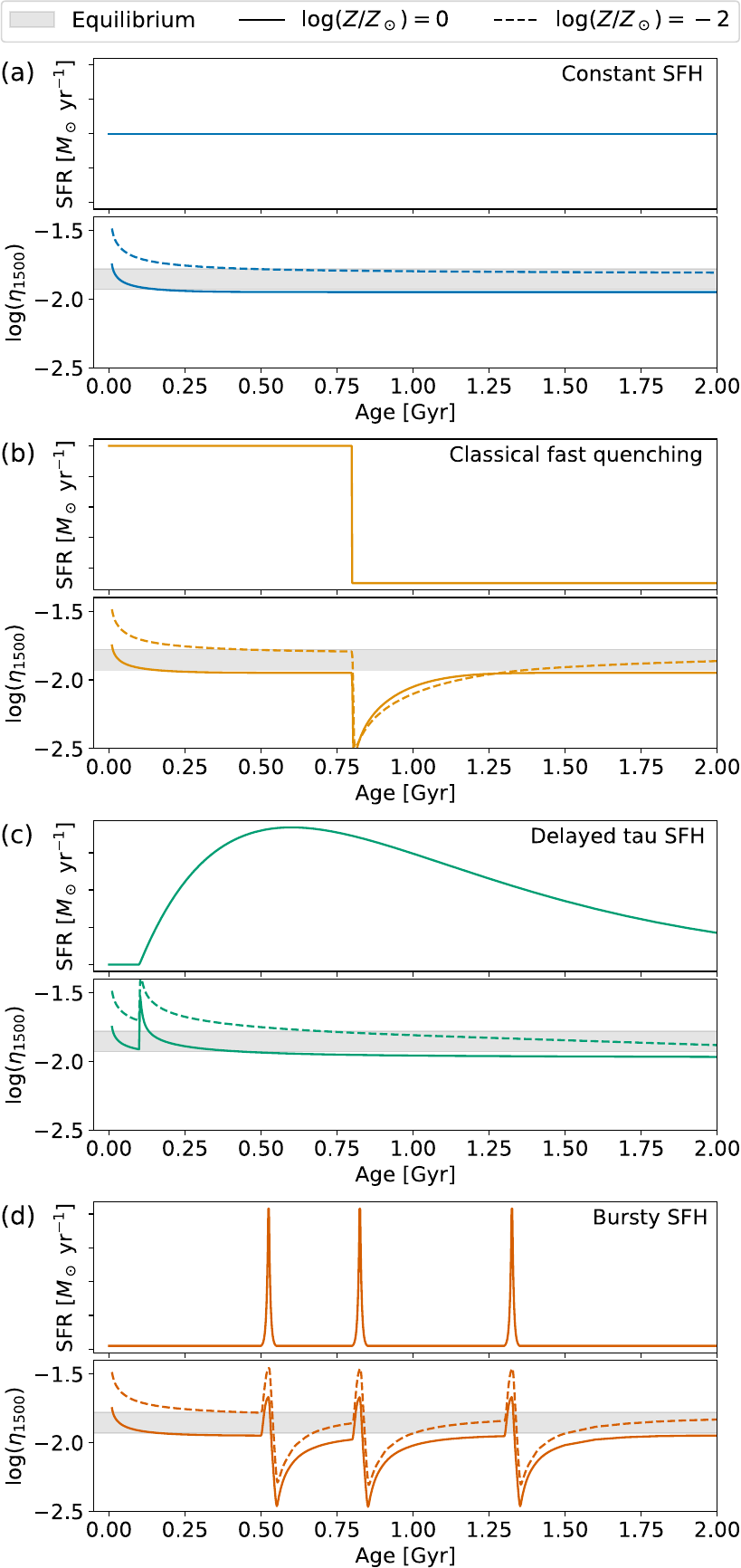}
    \caption{Illustration of how the burstiness parameter $\eta_{1500}$ behaves with four representative SFHs.
    In each panel from (a) to (d), the top sub-panel show the assumed SFH, and the bottom sub-panel presents the corresponding $\eta_{1500}$ evolution assuming solar (solid lines) or 1\%-solar metallicity (dashed lines).
    The horizontal shaded region denotes the possible range of $\eta_{1500}$ values for smooth and steady SFHs ($1/85<\eta_{1500}<1/60$; see the text for the details).
}
    \label{fig:eta_examples}
\end{figure}

The H$\alpha$-to-UV ratio has been used to quantitatively evaluate SFH burstiness for both local and high-$z$ galaxies \citep[e.g.,][]{Iglesias2004,Meurer2009,Weisz2012,Guo2016,Emami2019,faisst_recent_2019}.
Both the H$\alpha$ and far-UV luminosities of star forming galaxies are known as tracers of star formation, but they trace star formation on different timescales.
H$\alpha$ emission lines emerge from H{\sc ii} regions, which are the byproduct of ionizing radiation from short-lived stars such as late O- and early B-type stars, whereas the far-UV photons are produced also by longer-lived stars such as late B- and early-A type stars.
Consequently, the H$\alpha$ emission line and far-UV luminosity typically trace star formation activity on $\sim5$ Myrs and $\sim100$ Myrs timescales, respectively \citep[see also][and references therein]{kennicutt_star_1998}.
Therefore, if the SFR  is smooth and steady for a long time ($\gtrsim100$ Myr), the ratio of H$\alpha$ to UV luminosity should be approximately constant at an equilibrium value.
On the contrary, if the SFR fluctuates on short timescales ($<100$ Myr), the H$\alpha$ to UV ratio deviates from the equilibrium value: the ratio is higher when the SFR is rapidly rising, and is lower when the SFR is quickly declining.
To quantify star formation rate burstiness, we make use of the different timescales for the two SFR indicators and define an observable burstiness quantity $\eta_{\rm 1500}$ as
\begin{equation}\label{eqn:eta}
\eta_{\rm 1500}=\frac{F_{\rm H\alpha}}{\nu f_{\nu,1500}},
\end{equation}
where $F_{\rm H\alpha}$ is the H$\alpha$ line flux and $\nu f_{\nu,1500}$ is the monochromatic flux at rest-frame $\lambda_{\rm rest}=1500$ \AA. 
Here both the H$\alpha$ and UV fluxes are corrected for dust attenuation.
The wavelength at which the rest-frame UV flux is measured  (1500\AA  \ in this work)  is important to note since the numerical values of $\eta$ depend on it  (e.g., for a related but slightly different definition of the $\eta$ parameter see Estrada-Carpenter et al., in prep, who measure rest-frame UV flux at rest-frame 2300\AA).  However, the difference of the adopted rest-frame UV wavelength does not change the general principle of the H$\alpha$-to-UV burstiness parameter, even if it changes the numerical values of $\eta$.

Figure \ref{fig:eta_examples} illustrates how the burstiness parameter $\eta_{1500}$ behaves under four illustrative SFHs, calculated with Flexible Stellar Population Synthesis library \citep[FSPS;][see also Appendix \ref{apx:toy_model}]{Conroy2009}.
In each panel, top sub-panel shows the assumed SFH, and the bottom sub-panel shows how $\eta_{1500}$ evolves under the SFH, with assuming the matellicity of $\log(Z/Z_\odot)=0$ (solid lines) and $\log(Z/Z_\odot)=-2$ (dashed lines).
When a constant SFH is assumed (Figure \ref{fig:eta_examples} panel a), the $\eta_{1500}$ ratio stays in equilibrium state, but the equilibrium value may change with various assumptions such as stellar metallicity or IMF.
\citet{Mehta2022} extensively examined the possible range of this equilibrium values, and found the $\eta_{1500}$ should be $1/85<\eta_{1500}<1/60$ for constant SFHs, as long as the metallicity range of $\log(Z/Z_\odot)=-2$ to $0$ is considered.
The shaded horizontal regions in the bottom sub-panels in Figure \ref{fig:eta_examples} present this equilibrium range.
The $\eta_{1500}$ values do not largely deviate from this equilibrium range when the SFR time evolution is smooth and has a long ($>100$ Myr) time scale.
For example, when a classical fast quenching SFH with constant star formation followed by a sudden quenching event is considered (Figure \ref{fig:eta_examples} panel b), $\eta_{1500}$ value can be significantly lower than the equilibrium range just after the sudden quenching, and the $\eta_{1500}$ value swings back to equilibrium again a few 100 Myr after the quenching event.
On the contrary, if a typical delayed-$\tau$ SFH with a long timescale is considered (Figure \ref{fig:eta_examples} panel c), $\eta_{1500}$ is significantly larger than equilibrium just after the onset of this episode, and stays within the equilibrium range while the SFR is gradually changing.
Alternatively, if a "bursty" SFH is assumed (Figure \ref{fig:eta_examples} panel d), the $\eta_{1500}$ value is larger than equilibrium while the SFR is rapidly rising, and $\eta_{1500}$ becomes significantly smaller in the subsequent fast quenching phase.
As can be seen in Figure \ref{fig:eta_examples}, the burstiness parameter $\eta_{1500}$ can deviate from the equilibrium range ($1/85<\eta_{1500}<1/60$) only if the galaxy experiences an abrupt change in the SFR very recently ($\lesssim100$ Myr), and thus $\eta_{1500}$ can be used as a proxy of the star formation burstiness.

\subsection{This study}

Statistical investigation of SFH burstiness in high-$z$ galaxies has been carried out even before the launch of JWST using the previously most sensitive near-infrared (NIR) imaging facility, {\it Spitzer}/IRAC.
Since the H$\alpha$ emission lines at high-$z$ are redshifted to $\lambda_{\rm obs}>2\ \mu$m and unfeasible with ground-based telescopes, imaging observations with the IRAC 3.6 or 4.5 $\mu$m filters have long been used to identify the H$\alpha$ emission lines at $4\leq z \leq 6$ through their photometric imprint in galaxies' broadband spectral energy distributions \citep[SEDs; e.g.,][]{faisst_recent_2019}.
However, due to the lack of sensitivity and angular resolution of IRAC observations, only massive galaxies have been investigated in previous studies and it has been impossible to investigate the SFH burstiness and its environmental dependence in high-$z$ low-mass galaxies.

Thanks to the deep and high-resolution imaging data from JWST/NIRCam, it is now possible to investigate SFH burstiness and the local environment of low-mass galaxies in the high-$z$ Universe.
The reddest filter on JWST/NIRCam is F444W with which the H$\alpha$ emission line can be observed at $z\sim5$-$6$, thus this is the highest redshift where we can measure SFH burstiness with NIRCam using the H$\alpha$ line.
In addition, the medium-band filters on NIRCam are a powerful tool to identify/measure the emission lines in the high-$z$ Universe \citep[see e.g.,][]{Withers2023}.
Using a medium-band filter such as F410M in a combination with F444W enables reliable measurement of emission line fluxes through the photometry.

In this paper we give the first insight into these science topics of star formation burstiness and its relation with galaxy local environments in high-$z$ low-mass galaxies, exploiting NIRCam imaging taken as a part of the Canadian NIRISS Unbiased Cluster Survey \citep[CANUCS; see][for the detailed survey design]{willott2022}.
CANUCS is a JWST cycle1 GTO program, which observes five different gravitational lensing cluster fields with NIRISS, NIRCam and NIRSpec.
In this paper we use data on the first field observed by CANUCS, MACS J0417.5-1154 (hereafter MACS0417).
The paper is structured as follows.
We begin with the description of data reduction in Section \ref{sec:data}, and then present how we select our sample of galaxies at $z \sim 5$-$6$ in Section \ref{sec:sample}.
Next we present the methodology to derive the properties of sample galaxies in Section \ref{sec:analysis} including the burstiness parameter $\eta_{1500}$ and the identification of interacting galaxies.
Finally we show the results in Section \ref{sec:results} and discuss them further  in Section \ref{sec:discussion}.
The main results in this paper are summerized in Section \ref{sec:conclusions}.
Throughout this paper we assume the $\Omega_{\rm M}=0.3$, $\Omega_\Lambda=0.7$, $H_0=70$~${\rm km \: s^{-1} \: Mpc^{-1}}$ cosmology, the \citet{Chabrier2003} IMF, and all magnitudes are quoted in the AB system.

\section{Data reduction}
\label{sec:data}
\subsection{JWST and HST imaging observations}
This work utilized the NIRCam and NIRISS imaging observations obtained as a part of  CANUCS.
The survey observes the cluster center field in the NIRCam filters F090W, F115W, F150W, F200W, F277W, F256W, F410M, and F444W with exposure times of 6.4\,ks each, and also in NIRISS filters of F115W, F150W, and F200W with exposure times of 2.3\,ks each.
The 3$\sigma$ point source limiting magnitude ranges from 29.1 to 29.8 in these filters.
We also utilized archival data of HST/ACS imaging observations in this field with F435W, F606W, and F814W filters, and that of HST/WFC3 F105W, F125W, F140W, and F160W (HST-GO-11103 PI Ebeling, HST-GO-12009 PI von der Linden, HST-GO-14096 PI Coe, and HST-GO-16667 PI Brada\v{c}).

\begin{figure}
    \includegraphics[width=0.95\columnwidth]{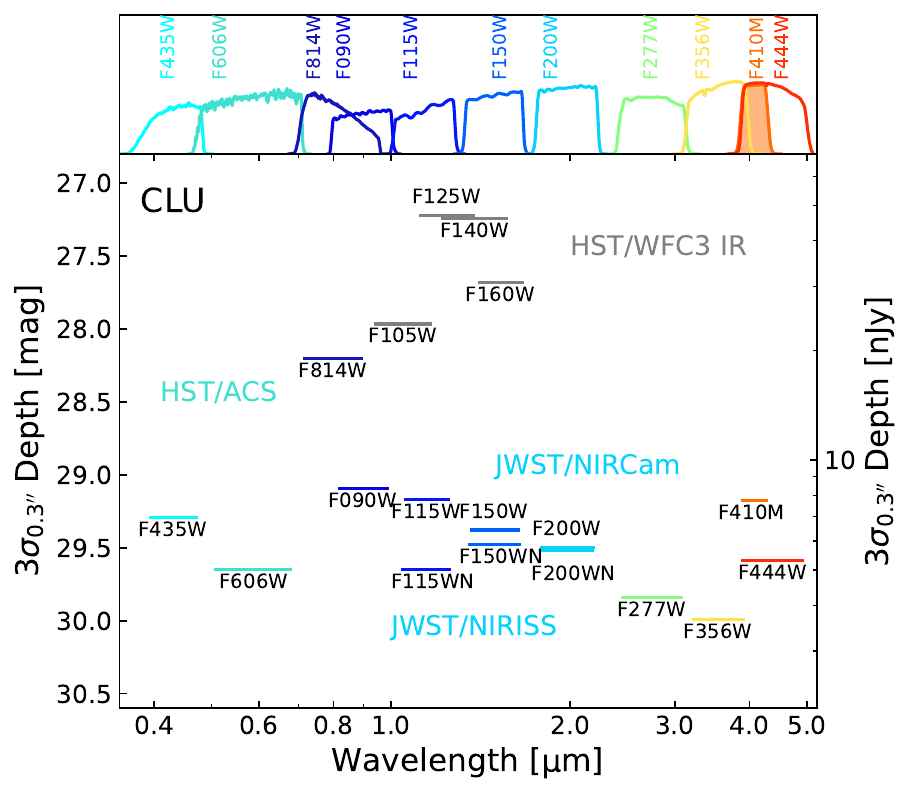}
    \includegraphics[width=0.95\columnwidth]{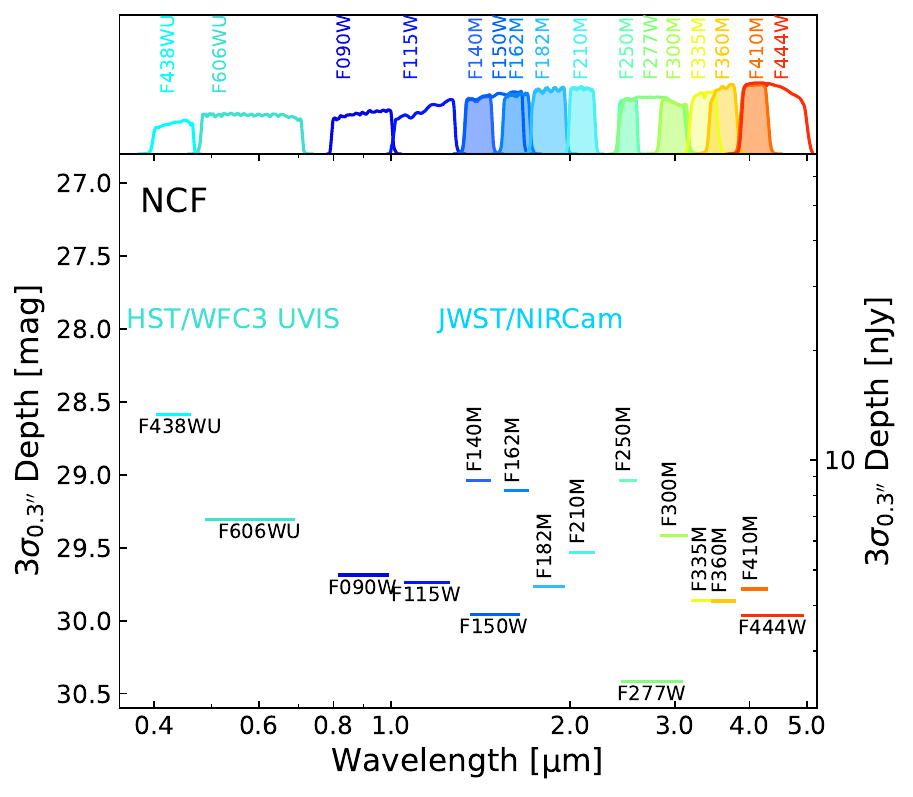}
    \caption{Depths of our imaging observations in the CLU field (top) and the NCF field (bottom).
    In each field, the $3\sigma$ depths measured in 0.3$^{\prime\prime}$-diameter aperture in all the available filters are shown.
    We also show the filter transmission curves at the top of each panel.
    Filled curves show the transmission curves of medium-band JWST/NIRCam filters, and open curves show those of broadband filters.
    For clarity, the filter transmission curves of HST/WFC3 IR and JWST/NIRISS filters are not shown.
    }\label{fig:filterset}
\end{figure}

In parallel with the NIRISS observation on the cluster center, CANUCS also observes a flanking field with 14 NIRCam filters:  F090W, F115W, F140M, F150W, F162M, F182M, F210M, F250M, F277W, F300M, F335W, F360M, F410M, and F444W with exposure times 10.3\,ks for 10 filters and 5.7\,ks for 4 filters.
The 3$\sigma$ point source limiting magnitudes in this flanking field are between 29.4 and 30.0 for most filters.
In the flanking field, we utilized archival HST imaging observations with WFC3/UVIS F438W and F606W (HST-GO-16667 PI Brada\v{c}).
We used both of the NIRCam + NIRISS imaging observations of cluster center field (referred as CLU hereafter) and the NIRCam observations of flanking field (referred as NCF hereafter).
Figure \ref{fig:filterset} summarizes the wavelength coverage and depths of our imaging observations in both fields.

The basic data reduction process is similar to that described in \citet{Noirot2023}.
We processed the NIRCam and NIRISS direct images with a combination of the official STScI JWST pipeline (version 1.8.0 and CRDS context {\tt jwst\_1001.pmap}) and {\tt grizli} version 1.6.0.
We then drizzled JWST and HST images on the same pixel scale of 40 miliarcsec/pixel, and assigned Gaia DR3 astrometry.
In CLU field images, we model and subtract the intracluster light and the brightest cluster galaxies (BCGs) using a custom code (Martis et al., in preparation) in each filter.
We also obtained the empirical point spread functions (PSFs) for all images from isolated stars in the field, and convolved the images with  kernels to match the PSFs in all images to that of the lowest resolution image, F444W (G.~Sarrouh et al., in prep.).

\subsection{Photometry}\label{subsec:photometry}
We created photometric catalogs separately in CLU and NCF field.
In each field, we first made a $\chi_{\rm mean}$-detection image \citep{Drlica-Wagner2018} combining all the available images from HST/ACS, WFC3, JWST/NIRCam and NIRISS in CLU, and from HST/WFC3 and JWST/NIRCam in NCF.
In this process we used the BCG-subtracted images before the PSF-convolution in order to maximize the signal-to-noise ratio (S/N).
The $\chi_{\rm mean}$-detection image is generally designed to optimally co-add all images and enable us to identify sources that are detected in any of the co-added images.
We  refer the reader to \citet{Drlica-Wagner2018} for further details.
In brief, for each pixel, the $\chi_{\rm mean}$ signal is calculated as
\begin{equation}
    \chi_{\rm mean} = \frac{\sqrt{\sum\limits_{i\leq n} f_i^2/\sigma_i^2} - m}{\sqrt{n-m^2}},
\end{equation}
where subscript $i$ refers to each filter used to generate the $\chi_{\rm mean}$ image, $n$ is the number of filters used, $f_{i}$ is the flux in the $i$th filter image and $\sigma_i$ is the associated error, and $m$ is desined as
\begin{equation}
    m = \sqrt{2}\frac{\Gamma((n+1)/2)}{\Gamma(n/2)},
\end{equation}
using the Gamma function $\Gamma(x)$.
We generated a $\chi_{\rm mean}$-map in each of CLU and NCF field following this procedure and using all available filters.

We performed source detection and photometry with the {\tt photutils} package \citep[version 1.5.0;][]{Bradley2022zndo...6825092B} in CLU and NCF.
We ran the source detection in each field using the $\chi_{\rm mean}$ detection image with the two-mode "hot+cold" detection strategy, and made the photometry in all F444W-PSF-convolved images with the same strategy as is usually done in dual-image mode with {\tt SExtractor} \citep[][]{Bertin1996}.
We put 0.3"-diameter circular apertures at the position of each detected source.
The aperture photometry was corrected for  Galactic extinction using $E(B-V)=0.03$ mag obtained from \citet{schlafly_measuring_2011} at the position of the MACS0417 field assuming the \citet{fitzpatrick_correcting_1999} dust-attenuation law with $R_V=3.1$.
Photometric uncertainties were estimated by putting 2000 empty apertures on the noise-normalized image in each filter.
The errors were then measured from the 1$\sigma$ width of the resulting Gaussian distribution of the empty-aperture fluxes multiplied by the noise level at the position of the source.
In total, we obtained 18249 (12556) detections in the CLU (NCF) field, and measured the fluxes of them in all available filter images.

We also used photometric redshift estimations in the following analyses  (Section \ref{sec:sample} and \ref{sec:analysis}), obtaining using the {\tt EAzY} code \citep{Brammer2008}.
We used the latest standard templates ({\tt tweak\_fsps\_QSF\_12\_v3}) supplemented with an additional template set by \citet{Larson2022}.
The latter templates were added as they are young ($\sim1$-$10$ Myr old), blue galaxy templates with strong rest-frame optical emission lines which have been shown to give proper fits to high-$z$ star-forming galaxies.
No zero-point corrections were applied, but a systematic error of 5\% of the flux in each filter was added in quadrature to the error budget.
In order to reduce the possibility of unphysical solutions, we also applied a magnitude prior, as is commonly done for photometric redshift measurements \citep[e.g.,][]{Brammer2008,Weaver2022,Desprez2023}.

\section{Sample selection}\label{sec:sample}
\begin{figure*}
	\includegraphics[width=2.0\columnwidth]{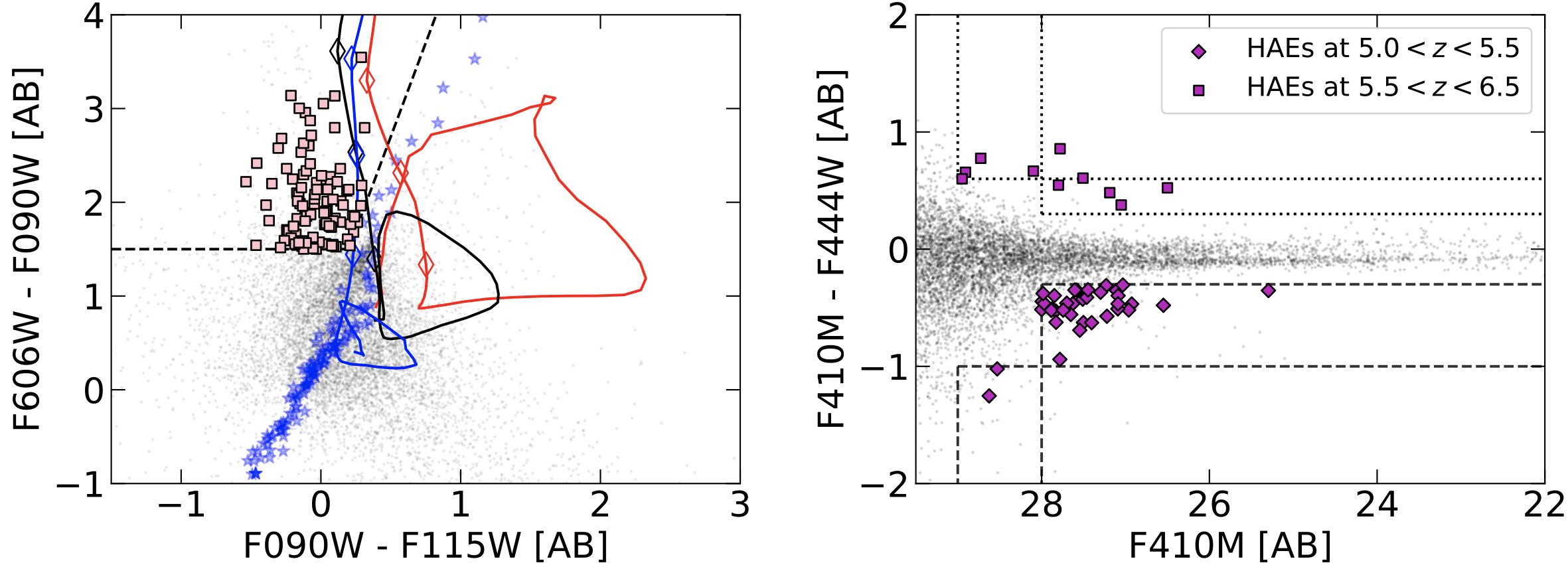}
    \caption{Sample selection criteria used in this work.
    \textit{Left}: F606W dropout selection using the colour-colour diagram of F606W-F090W vs. F090W-F115W.
    Gray points represent the full catalog both in CLU and NCF fields, and pink squares represent the LBG candidates that meet the selection criteria in Section \ref{subsec:LBG_sel}.
    The colour-colour window is shown with black dashed lines.
    Blue stars show the colours of local dwarf stars from \citet{Pickles1998} templates.
    Blue, black, and red solid curves present the evolutionary tracks of galaxy templates from \citet[Im, Sbc, and E, respectively]{coleman_colors_1980}.
    Coloured open diamonds show the colours of corresponding galaxy templates at $z=4.5$, 5.0, and 5.5.
    \textit{Right}: colour excess selection using the colour-magnitude diagram of F410M vs. F444W.
    Gray points represent the full catalog both in the CLU and NCF fields, and magenta diamonds (squares) represent the HAE candidates selected as F410M-excess sources in Section \ref{subsubsec:F410M_sel} (F444W-excess soursec in Section \ref{subsubsec:F444W_sel}).
    Dashed (dotted) lines show the colour criteria for F410M-excess (F444W-excess) sources.
}
    \label{fig:sample_selection}
\end{figure*}
In this section, we describe the sample selection strategy in this work.
Because we aim at investigating the star formation burstiness of galaxies at $z\sim5$-$6$, we need to construct a sample with various types of SFHs (i.e., wide range of the burstiness parameter $\eta_{1500}$).
One of the common ways to make a high-$z$ galaxy sample is by a simple photometric redshift selection.
Photometric redshift selections often require small photometric redshift uncertainties in order to remove spurious sources including low-$z$ interlopers \citep[e.g.,][]{faisst_recent_2019}.
However, photometric redshift uncertainties are expected to be smaller in extreme emission line galaxies that have high equivalent width lines as compared to more ordinary galaxies without strong  emission lines in their SEDs.
Therefore, a photometric redshift uncertainty cut can bias our sample towards the high-$\eta_{1500}$ population and missing low-$\eta_{1500}$ galaxies.
Thus, in this work, we made a sample of galaxies at $z\sim5$-$6$ based on two different complementary selection criteria primarily based on their rest-frame UV colours (Sec.~\ref{subsec:LBG_sel}) or H$\alpha$ emission lines (Sec.~\ref{subsec:med-excess_sel}).
We used the same colour-based sample selection criteria both in CLU and NCF.

\subsection{F606W-dropout selection}\label{subsec:LBG_sel}
Our first selection category isolates Lyman break galaxies (LBGs) at this redshift.
The Lyman break selection is a well-known technique to select high-$z$ galaxies based only on photometry, which uses the sharp break at $\lambda_{\rm rest}=1216$ \AA\ in the observed SED due to intergalactic medium (IGM) absorption to identify galaxies at a cosmic distance \citep[e.g., ][]{Steidel1996}.
At the redshift of $z\sim5$ - $6$, the Lyman break appears at $\lambda_{\rm obs}\sim7300$-$8500$ \AA, and we can identify these galaxies as F606W dropout.

To select LBGs at this redshift, we set the criteria in the colour-colour diagram of F606W, F090W, and F115W as follows (Figure \ref{fig:sample_selection} left):
\begin{gather}
    m_{606} - m_{090} > 1.5,\\
    m_{606} - m_{090} > 4(m_{090}-m_{115}) + 0.7.
\end{gather}
The colour criteria were determined by referring to the colour evolution of galaxy spectral templates from \citet[coloured solid lines in Figure \ref{fig:sample_selection} left]{coleman_colors_1980} with  IGM attenuation  following the prescription by \citet{madau_radiative_1995}, while also referring to the locus of local dwarf stars in this colour-colour diagram constructed from \citet{Pickles1998} templates (blue stars in Figure \ref{fig:sample_selection} left).
This selection is expected to be effective at $4.7\lesssim z \lesssim 6.1$.
In addition, we also required S/N both in F090W and F115W larger than 5 and S/N in F435W (F438W) smaller than 2 in the CLU (NCF) field.
We then removed any source whose segmented area in the detection map is spuriously small (i.e., less than 5 pixels) or whose position is too close to bad pixel regions or edge of the field-of-view (i.e., closer than 3 arcsec).
Finally we did visual inspections of all the candidates and removed spurious sources such as diffraction spikes or hot pixels.
We obtained a total of 78 (98) LBG candidates in the CLU (NCF) field.

However, among the 78 (98) candidates, we found a non-negligible number of apparent low-$z$ interlopers.
These sources are characterized with a weak dropout in F606W, quite red continuum colour in the NIRCam SW filters (i.e., F090W to F200W), and quite blue continua in the NIRCam LW filters (i.e., F277W to F444W).
Some of these sources also show diffuse morphology.
The SEDs of these sources are well explained with dusty star-forming galaxies at $z\sim0.5$, where we know the foreground galaxy cluster of MACS0417 is located ($z=0.443$), but poorly explained with galaxies at $z\sim5$-$6$.
Such sources are more likely to be found in the CLU field than NCF field ($\sim50$\ \% in the CLU sample while $\sim30$\ \% in the NCF sample).
Thus these sources could be (ultra-faint) foreground cluster members, but it is difficult to distinguish between this foreground contamination and high-$z$ galaxies only in the colour-colour diagram.
Thus, we used phot-$z$ estimates to remove these low-$z$ interlopers from the candidates.
We removed sources whose maximum likelihood redshifts from {\tt EAzY} ($z_{\rm ml}$) are smaller than $z=1.5$.
In the end, we obtained 38 (66) LBG candidates in total in CLU (NCF) field.
Pink squares in Figure \ref{fig:sample_selection} left show the resulting 104 LBG candidates.

\subsection{Medium-band colour selection}\label{subsec:med-excess_sel}
Although Lyman break selection is expected to produce a representative sample of (star-forming) galaxies at $z\sim5$-$6$, the sample can be biased towards rest-frame UV bright sources and can miss the UV-faint population.
Particularly in this study, we are keen to investigate  bursty SFHs among high-$z$ galaxies, but galaxies in the rising phase of a burst are expected to be relatively faint in rest-frame UV wavelength due to their high $\eta_{1500}$ values, thus we need to correct for this potential bias.
Galaxies experiencing a rapid SFR rise are expected to show strong H$\alpha$ line emission, and NIRCam medium-band colour selection has been shown to be successful in selecting such galaxy population with faint continuum and strong emission lines \citep[e.g.,][]{Withers2023}. Thus, we complement our sample with medium-band colour selection.
Namely, the H$\alpha$ emission line falls into both F410M and F444W ($5.0\leq z\leq5.5$) or only in F444W but not in F410M ($5.5\leq z\leq6.5$).
We therefore can select strong emission line galaxies at $z\sim5$ - $6.5$ by their extreme F410M - F444W colours.

\subsubsection{F410M-excess}\label{subsubsec:F410M_sel}
The H$\alpha$ emission lines from galaxies at $5.0\leq z\leq5.5$ boost both of F410M and F444W, but the excess in F410M is larger than in F444W due to the narrower bandwidth of F410M.
Thus these galaxies show particularly blue colour of F410M - F444W.
To select these sources, we set the following criteria in the colour-magnitude diagram of F410M versus F444W (dashed lines in Figure \ref{fig:sample_selection} right):
\begin{gather}
    m_{410} < 28\ \wedge\ m_{410} - m_{444} < -0.3, \label{eqn:f410m-1}\\
    {\rm or} \notag\\
    m_{410} < 29\ \wedge\ m_{410} - m_{444} < -1.0 \label{eqn:f410m-2}.
\end{gather}
In principle, this medium-band colour selection does not rely on rest-frame UV observations, and we can select sources regardless of whether a source is located in the field-of-view (FoV) of ACS/F606W (or WFC3/F606W in NCF).
However, in this work, to verify that the selected source is not a low-$z$ interloper and to homogenize the areas of LBG selection (Section \ref{subsec:LBG_sel}) and medium-band colour selection (Section \ref{subsec:med-excess_sel}), we only used sources located in the F606W FoV, and removed sources whose S/N in F606W is larger than 3.
In addition, similarly to Section \ref{subsec:LBG_sel}, we removed sources close to bad pixel regions and did visual inspections.
A total of 35 (29) F410M-excess candidates were obtained in the CLU (NCF) field.

The 35 (29) candidates include not only H$\alpha$ emitters at $z\sim5$ but also emission line galaxies at other redshifts.
Particularly [{\sc Oiii}]+H$\beta$ emitters at $z\sim7$ are also selected with this colour excess selection, because the [{\sc Oiii}]+H$\beta$ emission lines at $6.8 \leq z \leq7.6$ also boost F410M and F444W photometry.
To obtain a sample containing only H$\alpha$ emitters at $z\sim5$, we utilized photo-$z$ estimations from {\tt EAzY} (done in Section \ref{subsec:LBG_sel}), and selected sources with $4< z_{\rm ml} < 6$.
We obtained 17 (19)  H$\alpha$ emitters (HAEs) at $z\sim5$ with this F410M-excess selection in the CLU (NCF) field.
Magenta diamonds in Figure \ref{fig:sample_selection} right show these final 36 HAE candidates at $z\sim5$.

\subsubsection{F444W-excess}\label{subsubsec:F444W_sel}
In contrast to $5.0\leq z\leq5.5$, H$\alpha$ lines at $5.5\leq z \leq6.5$ boost only F444W but not F410M, which results in quite red  F410M - F444W colour.
We thus use the same colour-magnitude diagram of F410M versus F444W, and set criteria similar to Equation (\ref{eqn:f410m-1}) and (\ref{eqn:f410m-2}) but reversing the number signs (dotted lines in Figure \ref{fig:sample_selection} right):
\begin{gather}
    m_{410} < 28\ \wedge\ m_{410} - m_{444} > 0.3, \label{eqn:f444w-1}\\
    {\rm or} \notag\\
    m_{410} < 29\ \wedge\ m_{410} - m_{444} > 0.6 \label{eqn:f444w-2}.
\end{gather}
We followed the same additional procedures as Section \ref{subsubsec:F410M_sel}, and obtained 3 (10) F444W-excess candidates in CLU (NCF).

This F444W-excess selection can pick up emission line galaxies at other redshifts as well.
Similarly to the F410M-excess selection, [{\sc Oiii}]+H$\beta$ emission lines at $7.8 \leq z \leq9.0$ can result in similarly red F410M-F444W colour.
We thus again used phot-$z$ estimations and selected sources with $5<z_{\rm ml}<7$.
Following this, we finally obtained 2 (8) HAEs at $z\sim6$ with this F444W-excess selection in the CLU (NCF) field.
Magenta squares in Figure \ref{fig:sample_selection} right show the final 10 HAE candidates at $z\sim6$.

\begin{figure}
	\includegraphics[width=\columnwidth]{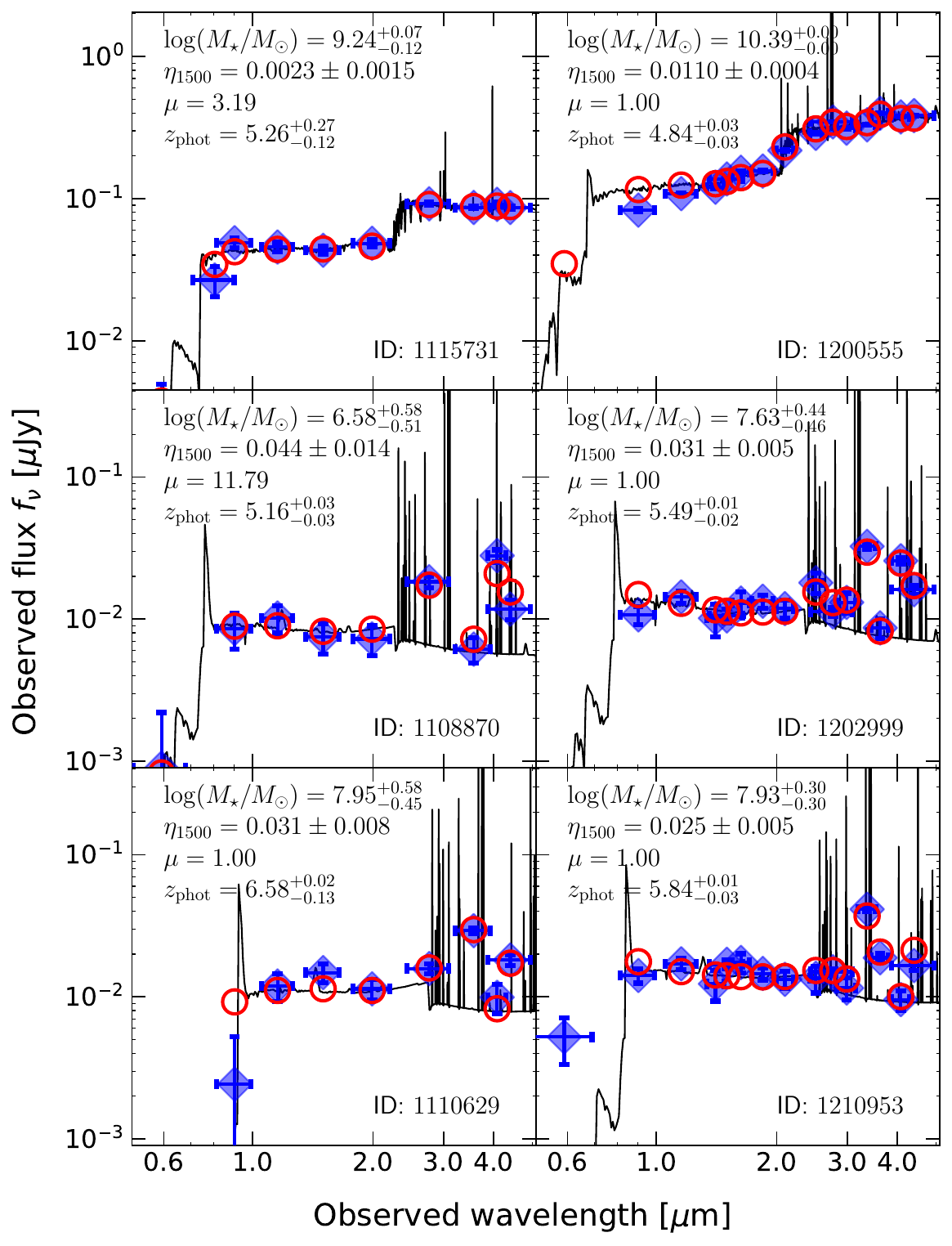}
    \caption{Example SEDs of galaxies in the sample of this work.
    Blue filled diamonds with error bars show the observed SED, and the black solid line is the best-fitted galaxy spectrum obtained in Section \ref{subsec:SEDfit}.
    Red open circles show the synthesized model photometry of the best-fitted spectrum. In this plot, photometry in WFC3/IR and NIRISS filters are not displayed.
    Galaxies in the left column are located in the CLU field, and those in the right column are in the NCF field.
    From the top to bottom, examples that meet selection criteria only of LBG, F410M-excess, and F444W-excess selection are shown, respectively.
    In each panel, we also insert their physical properties including the stellar mass, $\eta_{1500}$ ratio, magnification factor, and the photometric redshift (see text for the details).
    The stellar masses are corrected for the lens magnification.
    }
    \label{fig:SEDs}
\end{figure}

\subsection{Our final combined sample}
With the criteria above, we obtained 38 (66) LBG candidates and 19 (27) HAE candidates in the CLU (NCF) field.
We took the union of these three set of candidates in each field and combined the two fields, resulting in a sample of 125 galaxies at $4.7\lesssim z \lesssim 6.5$.
Note that some of the sample galaxies are identified both as LBGs and HAEs.
In Figure \ref{fig:SEDs}, we show example SEDs of different types of objects in our sample.
The left column presents examples in the CLU field and the right column presents those in the NCF field, thus the filter sets in left and right columns are different.
From top to bottom in the figure, we show example galaxies that are selected only through LBG selection (Section \ref{subsec:LBG_sel}), F410M-excess (Section \ref{subsubsec:F410M_sel}), and F444W-excess (Section \ref{subsubsec:F444W_sel}) selection, respectively.
It can be seen that the LBG selection is able to select UV bright galaxies (not necessarily with strong emission lines), while the medium-band colour selected galaxies show strong excesses in their photometry due to the extreme emission lines.

\section{Analysis}\label{sec:analysis}
\subsection{Lens modeling in the CLU field}\label{subsec:lens}
Galaxies in our CLU field can be strongly magnified by the gravitational lensing due to the foreground galaxy cluster. To account for this, we used the gravitational lens model by Desprez et al. (in prep.), to estimate the magnification factors and source-plane positions for each CLU  galaxy in our sample.
In contrast to the CLU field, the gravitational lensing effect in the NCF field is small enough (median $\mu\sim1.1$) and consequently sample galaxies in the NCF field were neglected in this lensing analysis.

Our lens model is built using \texttt{Lenstool} \citep{Kneib1993,Jullo2007}.
The model leverages the multiple image constraints from \cite{Mahler2019} and \cite{Jauzac2019}, as well as new multiple image systems and redshifts we obtained from the CANUCS-\textit{JWST} data.
The model includes cluster-size- and galaxy-size halos described as double Pseudo-Isothermal Elliptical (dPIE) profiles \citep{Eliasdottir2007arXiv}.
The details of this updated model are to be presented in Desprez et al. in prep.

With the updated lens model, we used {\tt Lenstool} to obtain the magnification factor and the source-plane position of each sample galaxy, and also to predict the positions of their multiple images and searched for multiply-imaged systems in our sample.
As a result of this, we found  that two sources in the sample are doubly imaged and both of the double images were included in our sample catalog.
This doubly-imaged system is a close pair of two ultra-low-mass ($M_\star\sim 10^7 M_\odot$) galaxies, first reported and dubbed as MACS0417-ELG1 and -ELG2 by \citet{Asada2023}. The two galaxies are doubly imaged for a total of four apparently different galaxies, all of which are included in our sample.
To avoid double-counting in our statistical analysis,  we removed one of the double images of the galaxy pair and used only one of its double images.
This reduced the number of sample galaxies to 123.

\subsection{Properties from SED fitting}\label{subsec:SEDfit}
We ran the SED fitting code {\tt dense basis} \citep{Iyer+17,Iyer+19} to estimate the physical properties including stellar mass, star-formation rate (SFR), dust attenuation value ($A_V$), (gas-phase) metallicity, and non-parametric star-formation histories (SFHs).
We assumed the  \citet{Chabrier2003} IMF and \citet{Calzetti2000} law for dust attenuation.
{\tt Dense basis} uses a Bayesian approach in the fitting, which requires us to assume the prior distributions of these physical parameters.
Since the default prior configurations are optimized to lower redshift galaxies than our sample, we adjusted several physical parameter priors to high-$z$ galaxies:
we assumed a flat sSFR prior between $-3 < \log({\rm sSFR}/{\rm Gyr}^{-1}) < 2.5$, flat metallicity prior between $-2.5 < \log(Z/Z_\odot) < -1.0$, and an exponential $A_V$ prior with the scale value of $A_V=3.0$ mag.
We also allowed for more bursty SFHs than the default configuation by varying the Dirichlet concentration parameter that sets the SFH prior\footnote{Note that {\tt dense basis} does not assume a SFH but infers the non-parametric SFHs from the data.}.
The redshift was allowed to vary around the best phot-$z$ estimation ($z_{\rm ml}$) from {\tt EAzY} by $\Delta z=\pm0.5$\footnote{The typical photo-$z$ uncertainty of sample galaxies from {\tt EAzY} is $\sim0.04$, and the window is large enough to cover the photo-$z$ uncertainty.}.
We then obtained the best-fitting model spectra and best estimations of the physical parameters for each sample galaxy. 
Here we fit to the fluxes in HST/ACS (WFC3 UVIS) and JWST/NIRCam filters in CLU (NCF) field, and used aperture-corrected 0.3"-diameter aperture fluxes in SED fitting.
The aperture correction factors were estimated from the ratio of Kron to 0.3"-diameter aperture fluxes in the F444W image.
The physical quantities were then corrected for lens magnification for sources in the CLU field.

\begin{figure}
	\includegraphics[width=\columnwidth]{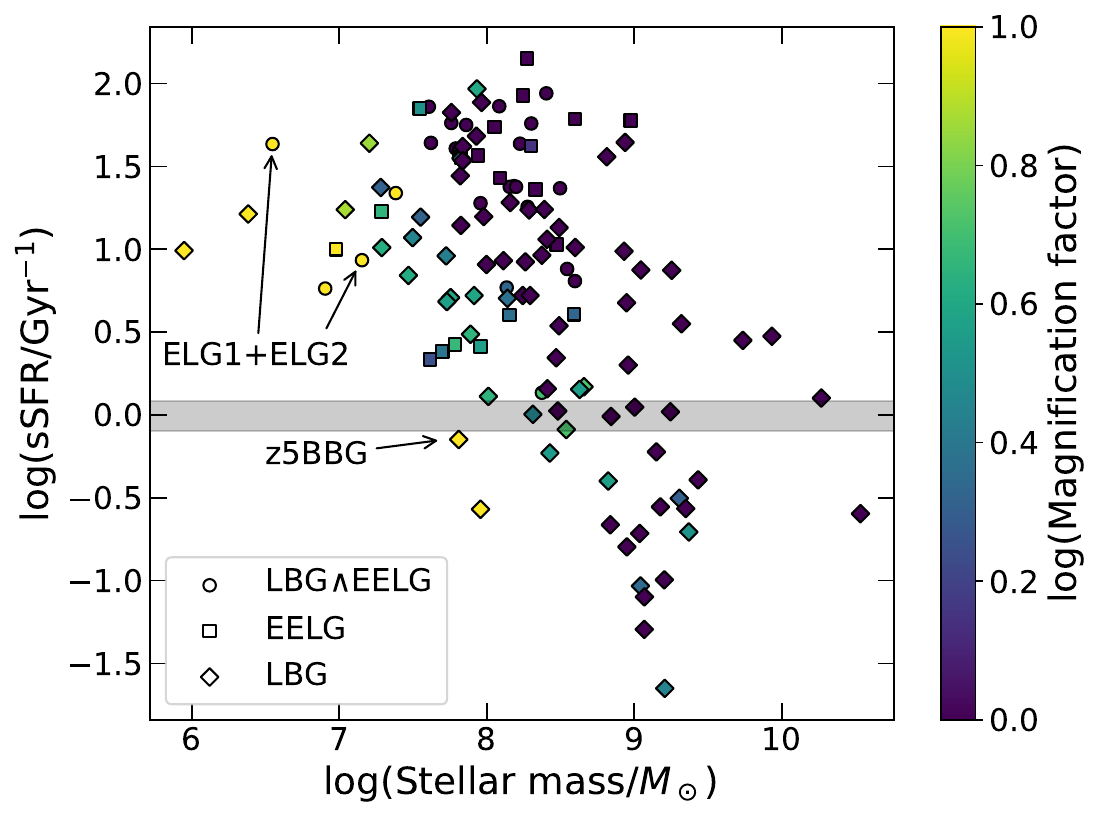}
    \caption{sSFR vs. stellar mass plot of the sample galaxies estimated by SED fitting in Section \ref{subsec:SEDfit}.
    Plots are colour-coded with the lensing magnification factor $\mu$, but $\mu$ is assumed to be unity for galaxies in the NCF field.
    The shape of symbols denotes which selection criteria the source meets (circle: both LBG and medium-band colour selection; square: only medium-band colour selection; diamond: only LBG).
    The gray shaded horizontal region presents the inverse of the age of Universe at $z=4.7$ - $6.5$, and galaxies above/below this region are often regarded as SF/quiescent galaxies.
    Among the quiescent galaxy candidates one source is spectroscopically confirmed to be a recently quenched galaxy \citep[MACS0417-z5BBG;][]{Strait2023}, and two of the SF galaxies are the ultra low-mass merging pair recently reported by \citet[MACS0417-ELG1 and -ELG2]{Asada2023}.
    }
    \label{fig:sSFR_mstar}
\end{figure}

Figure \ref{fig:sSFR_mstar} shows the stellar mass vs. sSFR distribution of the sample galaxies.
The gray shaded horizontal region in the figure represent the inverse of the age of Universe at the target redshift, and galaxies above/below this region is often regarded as SF/quiescent galaxy candidates at the redshift.
The sample is mainly composed of low-mass ($M_\star\sim10^8\ M_\odot$) star forming galaxies, but also includes some fraction of relatively quiescent galaxies.
One of the quiescent galaxy candidates in this plot is indeed spectroscopically confirmed to be a recently quenched galaxy at $z=5.20$ by \citet[MACS0417-z5BBG]{Strait2023}.
In addition, Figure \ref{fig:sSFR_mstar} demonstrates that the lowest-mass galaxy population ($M_\star\lesssim10^{7.5}\ M_\odot$) at this redshift can be effectively investigated with the aid of gravitational lensing effect. 

\subsection{Measurements of burstiness parameter $\eta_{1500}$}\label{subsec:Ha_meas}
As discussed in Section \ref{sec:intro}, the H$\alpha$-to-UV flux ratio $\eta_{1500}=F_{\rm H\alpha}/\nu f_{\nu, 1500}$ (Eq.~\ref{eqn:eta}) is a good indicator of the SFH burstiness in a galaxy.
To obtain the $\eta_{1500}$ value in each sample galaxy, we measured the H$\alpha$ and rest-frame UV flux in the following way.

In the H$\alpha$ measurements, for each sample galaxy, we first identified the broadband/medium-band filter where the H$\alpha$ emission line falls using the best-fitting phot-$z$ from {\tt EAzY}.
We then used the photometry in this filter for H$\alpha$ flux estimation in the sample galaxy.
If the H$\alpha$ line falls in both of F410M and F444W, we used F410M filter.
We then used the best-fitting spectrum from {\tt dense basis} obtained in Section \ref{subsec:SEDfit} to estimate the underlying continuum level in the filter.
Subtracting this continuum level from the observed photometry in the filter provided the excess due to emission lines including H$\alpha$, and we computed the H$\alpha$ flux from this excess measurement.
The H$\alpha$ flux uncertainties were estimated directly from the flux errors in the filter, and no uncertainties on the best-fitting spectrum (i.e., continuum level) was considered.
Dust attenuation corrections were applied using the $A_V$ values from {\tt dense basis} and \citet{Calzetti2000} attenuation law.

In the H$\alpha$ flux measurements above, we required a medium/broadband filter where the H$\alpha$ line falls.
However, based on the best-fitting phot-$z$ from {\tt EAzY}, the H$\alpha$ emission lines from part of the sample galaxies cannot be observed with the available filter set particularly in the NCF field.
The spectral coverage of NCF filter set has a gap at $\lambda\sim3.8-3.9\ \mu$m, and it is impossible to measure the H$\alpha$ flux from $4.8\leq z_{\rm ml}\leq4.9$ sources identified via LBG selection.
In addition, some other LBGs have high $z_{\rm ml}$ where H$\alpha$ moves out even from F444W.
We removed these galaxies only in the burstiness parameter $\eta_{1500}$ calculation, and do not use them in the following analyses that use the $\eta_{1500}$ values.

On the other hand, in the rest-frame UV monochromatic flux measurements, we directly used the photometry in a filter that  corresponds to $\lambda_{\rm rest}\sim1500$ \AA.
Namely, we used NIRCam/F090W photometry for sources at $z_{\rm ml}<5.57$ and NIRCam/F115W at $z_{\rm ml}>5.57$, and measured the (band-pass averaged) monochromatic flux.
Similar to the H$\alpha$ flux measurements, we corrected for the dust attenuation.
Using these measurements, we derived the (dust-corrected) $\eta_{1500}$ ratio in each sample galaxy.

Note that, in the H$\alpha$ measurements above, the excess in the H$\alpha$ filter was attributed only to the H$\alpha$ emission line and contributions from other emission lines were neglected.
In particular, [{\sc Nii}]$\lambda6583$ can contaminate photometry-based H$\alpha$ flux measurements.
For example, \citet{faisst_recent_2019} measured H$\alpha$ fluxes in high-mass ($M_\star \sim 10^{10}\ M_\odot$) galaxies at $z\sim5$ based on IRAC photometry, while assuming the [{\sc Nii}]/H$\alpha$ ratio to be 0.15.
However, recent JWST spectroscopic observations of low-mass, high-$z$ galaxies have reported non-detections of the [{\sc Nii}]$\lambda6583$ line or quite low [{\sc Nii}]$\lambda6583$/H$\alpha$ ratios in most cases \citep[e.g.,][]{Curti2023,Sanders2023,Nakajima2023} and the effect of [{\sc Nii}]$\lambda6583$ is expected to be small in our sample. 
This is consistent with the low [{\sc Nii}]$\lambda6583$/H$\alpha$ ratio predicted for metal poor galaxies \citep[e.g.,][]{Nakajima+22,Hirschmann2023}: given the typical stellar mass of our sample and the mass metallicity relation at this redshift \citep[][]{Curti2023}, the [{\sc Nii}]$\lambda6583$/H$\alpha$ ratio is expected to be $\lesssim0.05$.
We thus used H$\alpha$ flux measurements directly from the excess in the H$\alpha$ filter without correcting them for contributions from other emission lines such as [{\sc Nii}].
We also note that even if we assumed the high ratio of 0.15 as \citet{faisst_recent_2019}, the H$\alpha$ flux and resulting $\eta_{1500}$ ratio would systematically decrease by only 0.06 dex, which would not alter the results of this paper.

\subsubsection{Validation of photometric emission line flux measurements by comparison with spectroscopy}\label{subsubsec:spec_phot}
We obtained follow-up spectroscopic observations for 20 of our sample galaxies in this work.
Detailed analyses of their spectroscopic properties will be present in a future paper, but here we use the spectroscopy to validate our method of H$\alpha$ flux measurement from photometry.
The follow-up observation was acquired with JWST NIRSpec using PRISM/CLEAR disperser and filter, and the exposure time was $\sim3$ ks.
The spectroscopic data was processed in the same way as \citet{Withers2023}.
Having these processed spectrum, we used {\tt msaexp} to extract individual 1D spectra by combining the 2D spectra with an inverse-variance weighted kernel following \citet{Horne_1986}.

For the 20 spectroscopic sources, we identified the H$\alpha$ emission line and fitted single Gaussian + offset function to estimate their spectroscopic redshift ($z_{\rm spec}$) and H$\alpha$ equivalent widths (EWs).
We omitted 3 out of the 20 sources in this step because the H$\alpha$ falls the detector gap (1 source) or H$\alpha$ detected region is contaminated in the 2D spectrum (2 sources).
We first compared their $z_{\rm spec}$ and $z_{\rm phot}$ for the 17 spectroscopic sources, and found these two agree quite well ($\sigma_{\rm NMAD}=0.0079$).
There is only one catastrophic failure, whose photometric redshift was $z_{\rm ml} = 6.78$ but spec-z was revealed to be $z_{\rm spec}=5.5006$.
Other than this one exception, the spectroscopic redshift for the 16 spectra agree with our photometric redshift estimations, and the small scatter ($\sigma_{\rm NMAD}$) verifies our use of photometric redshift to identify the filter H$\alpha$ falls into.

\begin{figure}
	\includegraphics[width=\columnwidth]{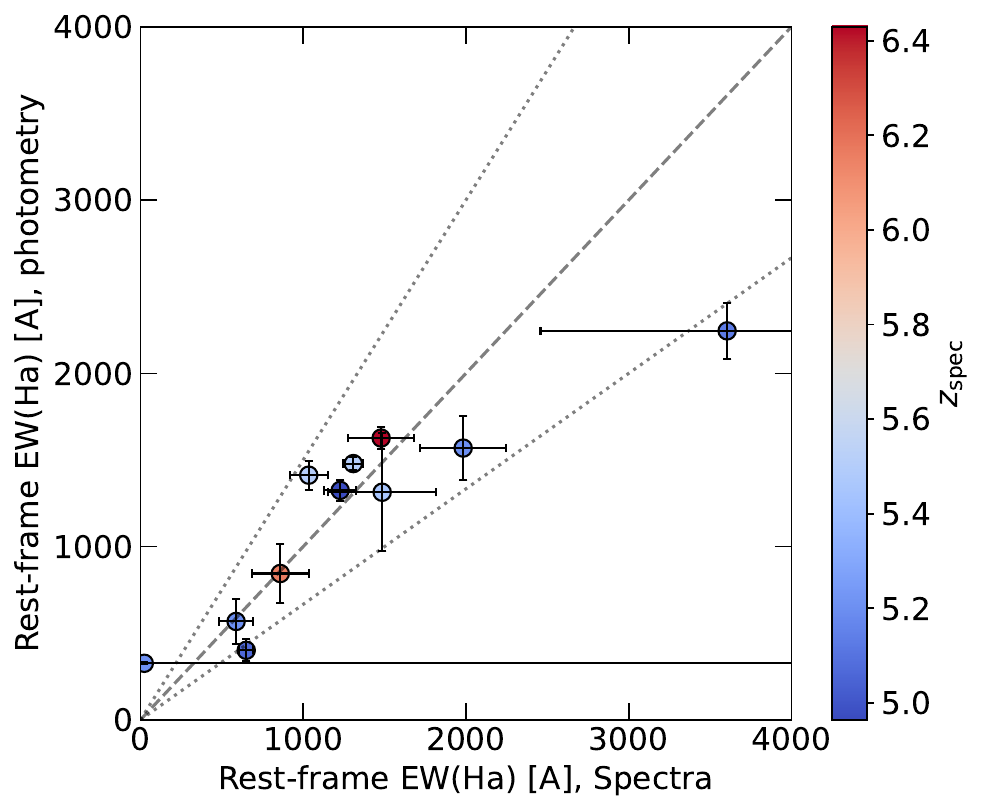}
    \caption{The observed H$\alpha$ EWs calculated from NIRSpec spectra vs. photometry with the methodology in this work (Section \ref{subsec:Ha_meas}).
    Plots are colour-coded with the $z_{\rm spec}$.
    The black dashed line denotes the one-to-one relation, and dotted lines present 1.5-to-1 relations.
}
    \label{fig:spec_phot}
\end{figure}

We next compared the spectroscopy-based H$\alpha$ EWs (EW$_{\rm spec}$) and photometry-based EWs (EW$_{\rm phot}$).
Here we further removed 6 spectroscopic sources because the underlying continuum is not detected in the spectrum (2 sources) or the photometric redshift is out of valid range for photometry-based H$\alpha$ measurement (4 sources; i.e., $z_{\rm ml}>6.5$ or $4.8<z_{\rm ml}<4.9$ in the NCF field).
Figure \ref{fig:spec_phot} shows the good agreement between the H$\alpha$ EW from spectra and photometry.
The NMAD of $({\rm EW_{phot}} - {\rm EW_{spec}})/{\rm EW_{spec}}$ is 0.176, and this is almost comparable to the typical uncertainty on our $\eta_{1500}$ calculation ($\Delta\eta_{1500}/\eta_{1500}\sim0.157$), and we could conclude that our methodology to estimate emission line fluxes (and resulting burstiness parameter $\eta_{1500}$) from photometry should be reliable within their uncertainties.

\subsection{Interacting galaxy identification}\label{subsec:morph_cls}
In this study, to evaluate the effect of the local environment on the burstiness of low-mass high-$z$ galaxies, particularly  galaxy-galaxy interactions, we also needed to identify those of our sample galaxies possibly experiencing galaxy interactions.
To this end, we used two different methodologies identifying possible galaxy-galaxy interactions in the sample.

\subsubsection{Separations from the nearest neighbor}\label{subsubsec:sep_NN}
One of the  simplest and easiest ways to identify interacting galaxies is by using  separation from the nearest neighbor (NN) in the sample.
Some apparent close pairs of galaxies in the sky could be chance projection and may not be physically interacting with each other.
However, given that the probability of finding (randomly distributed) projected sources at different redshift at a radius $r$ should increase in proportion to $r$, we can minimize the contamination from chance projected sources by setting a small enough separation threshold. We can expect galaxies closely paired with a smaller separation than the threshold to be at the same redshift and physically interacting with each other.

In this work we set the separation threshold to be $1^{\prime\prime}$ in the source plane, which at $z\sim 5.8$ corresponds to $\sim5.8$ physical kpc.  For each sample galaxy, we measured the separation from its nearest neighbor galaxy that is also in the sample, $r_{\rm NN}$. We stress that we did the $r_{\rm NN}$ measurement in the source plane, and we classified sample galaxies whose $r_{\rm NN}$ is smaller than $1^{\prime\prime}$ as interacting galaxies. Hereafter we refer to these galaxies as 'Class II'. 
We discuss the validity of this separation threshold and statistical evaluation of contamination from chance projections later in Section \ref{subsec:merger_frac}.

\subsubsection{Visual inspections}\label{subsubsec:VI}

Although the classification using $r_{\rm NN}$ in Section \ref{subsubsec:sep_NN} gives a subsample of interacting galaxies, visual inspections in the NIRCam images shows that some of the galaxies apparently experiencing galaxy-galaxy interactions in the sample are not classified as {\it interacting galaxies} (Class II) in Section \ref{subsubsec:sep_NN}  because their source-plane separations from the NN in the sample are greater than our separation threshold (i.e., $r_{\rm NN}>1^{\prime\prime}$).
The main reasons for these misclassifications are (1) sample incompleteness and (2) false deblending in during the source detection stage (Section \ref{subsec:photometry}).
The first case (1) can occur when the close companion to the sample galaxy is indeed detected and included in the catalog but does not pass the sample selection criteria due to e.g., its faintness.
The second case (2) mainly happens when the two interacting galaxies are bright and particularly close to each other, which results in a blended image in the detection map even with the high spatial resolution of JWST/NIRCam, and the source detection in Section \ref{subsec:photometry} failed to deblend the two galaxies (see e.g., ID 1200422 in Figure \ref{fig:postage_stamps_all_NCF1}).
To recover these misclassified galaxies into the interacting galaxy sample, we did visual inspection of every sample galaxy that was not classified as ClassII in Section \ref{subsubsec:sep_NN}.
We then reclassified a sample galaxy as interacting  if the galaxy met either of the two following criteria (hereafter we refer to these additional interacting galaxies as 'Class II*'):
(1) the galaxy has a close companion within $1^{\prime\prime}$  that is included in the photometric catalog and its best-fitting phot-$z$ from {\tt EAzY} is within the target redshift range; or (2) the source is blended and unsegmented in the detection image, the blended source seems clearly a different source, and the blended source also drops out in ACS/F606W (or WFC3/F606W in NCF).
In the following analysis, we use the supplemented sample of 'Class II'+'Class II*' as the fiducial sample of interacting galaxies, but use only 'Class II' sample as the most conservative sample of interacting galaxies.

For the rest of our sample galaxies, which are not classified either of Class II or Class II*, we further classified into two morphology classes based on the visual inspections: single, isolated sources (Class I) or turbulent/multi-component structures (Class III).
In the following, we basically do not separate Class I and Class III and treat both classes as galaxies without interaction otherwise specified.
In Appendix \ref{apx:postage_stamps_all}, we show the RGB cutout images of all the sample galaxies with their morphological Class classifications based on the visual inspections performed here.

In summary, our sample of 123 galaxies was classified into the following morphological classes: 
\begin{itemize}
    \item Class I (60 galaxies):   isolated galaxies with no sign of high-$z$ companions or complicated morphology,
    \item Class II and II* (23 + 24 = 47 galaxies):  galaxies with close companions at the similar redshift (within 1$^{\prime\prime}$ in the source plane, equivalent to $\sim$5.8 physical kpc at $z\sim5.8$),
    \item Class III (16 galaxies): galaxies with complicated morphologies that may or may not result from galaxy-galaxy interactions.
\end{itemize}
It is worth noting that Class II* and Class III can be hard to tell apart in some cases, particularly when the possible two galaxies are heavily blended.
However, the number of these galaxies are small and they do not affect the results in this paper:
part of such heavily blended sources are classified as Class II*, but results in the paper do not change even if we use only Class II as interacting galaxy sample.

\section{Results}\label{sec:results}
\subsection{Star formation burstiness and galaxy-galaxy interactions}\label{subsec:burstiness_and_interaction}
\subsubsection{Bursty SFHs in high-$z$ low-mass galaxies}\label{subsubsec:burstiness}

\begin{figure}
    \includegraphics[width=0.95\columnwidth]{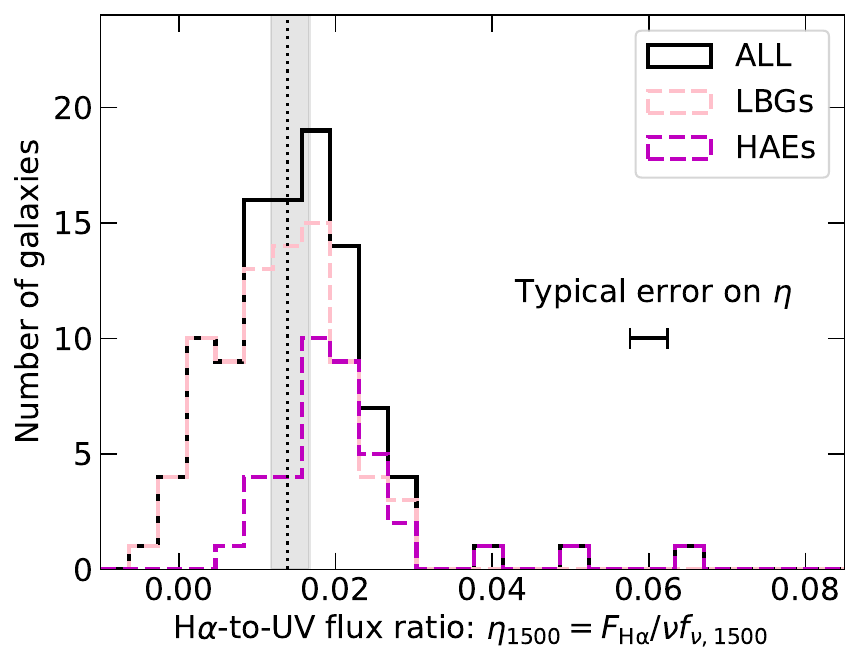}
    \caption {Histogram of $\eta_{1500}$ values.
    Black solid line shows the $\eta_{1500}$ histogram of our full sample, and the magenta (pink) dashed line shows that of sample galaxies identified by medium-band colour (LBG) selection.
    Typical uncertainty on $\eta_{1500}$ values is shown by the inset error bar.
    The gray shaded region shows the range of $\eta_{1500}$ values expected in the steady constant SFHs \citep[$1/85<\eta_{1500}<1/60$; e.g.,][]{Mehta2022}, and vertical black dotted line denotes the mean $\eta_{1500}$ value in the equilibrium state ($\eta_{1500}=1/72$).
    }
    \label{fig:eta_dist}
\end{figure}

Figure \ref{fig:eta_dist} shows the distribution of H$\alpha$-to-UV flux ratios $\eta_{1500}$ in the full sample.
The gray vertical band shows the equilibrum range predicted with steady SFHs \citep[$1/85 < \eta_{1500} < 1/60$; e.g., Figure \ref{fig:eta_examples}; ][, see also Appendix \ref{apx:toy_model}]{Mehta2022}
and our galaxies' $\eta_{1500}$ values are widely distributed from below to above this equilibrium range. 
Indeed, $60\%$ (63 of 105) of the sample galaxies have $\eta_{1500}$ values that deviate from the equilibrium range at $>1\sigma$ level (i.e., $\eta_{1500}+\Delta \eta_{1500}<1/85$ or $1/60<\eta_{1500}-\Delta \eta_{1500}$), thus the majority of the sample galaxies cannot be explained with steady and smooth SFHs.
This indicates that the low-mass galaxies at $4.7\lesssim z \lesssim 6.5$ in general are experiencing bursty SFHs rather than steady and smooth SFHs, and bursty SF should be the main growth channel in the high-$z$ low-mass galaxies.

In addition, Figure \ref{fig:eta_dist} shows that the combination of LBG and medium-band colour selection is required to cover the full range of $\eta_{1500}$ values.
The LBG selection produces a galaxy sample at our target redshift, but $\sim21\%$ (18 of 87) of LBGs have $\eta_{1500}$ values that are lower than the equilibrium range at $>1\sigma$ level.
These galaxies are thought to be in the declining phase of SFHs, and 
their SFRs predicted from the rest-frame UV luminosity using canonical conversion factors are relatively higher than their instantaneous SFRs.
On the other hand, $\sim27\%$ (12 of 44) of the high-$\eta_{1500}$ population (i.e., galaxies with $\eta_{1500}$ ratio higher than equilibrium range at $>1\sigma$ level) is selected only by medium-band colour selection, and LBG selection missed this population.
These facts suggest that there may be a systematic effect in cosmic star formation activity estimations (e.g., SFR density) that are usually based on rest-frame UV luminosity from Lyman-break selected samples.
We leave a further and quantitative investigation of this potential systematic effect to future work where a larger sample is available, and instead next turn to investigating the role of galaxy-galaxy interactions in contributing to the observed bursty SFHs in our sample.

\begin{figure}
	\includegraphics[width=\columnwidth]{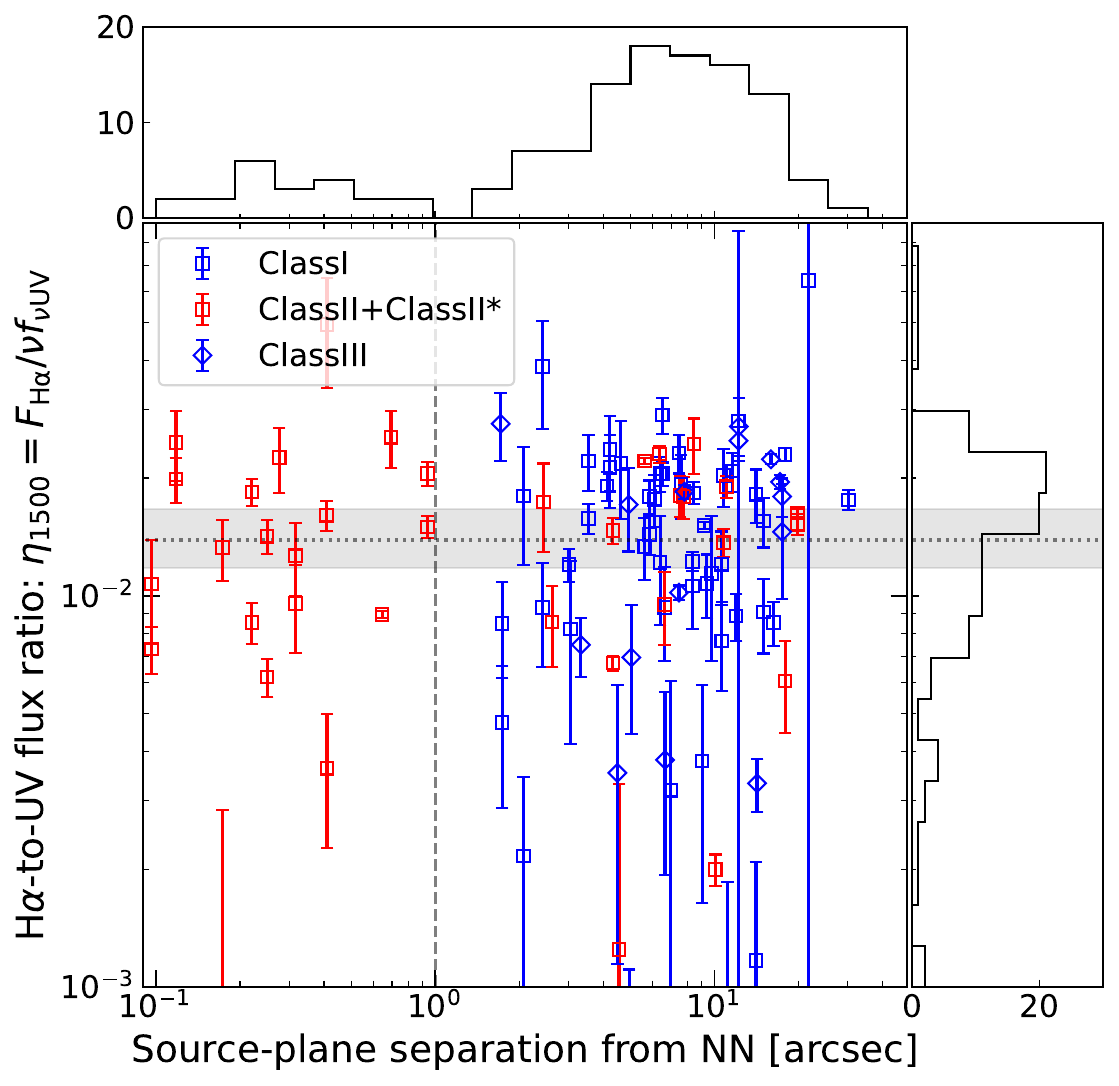}
    \caption{SFH burstiness ($\eta_{1500}$) vs. separation from the nearest neighbor in the sample ($r_{\rm NN}$).
    The top and right panels show the nominal histograms of $r_{\rm NN}$ and $\eta_{1500}$, respectively.
    Red points show sample galaxies classified a interacting in Section \ref{subsubsec:VI}; blue points mark non-interacting galaxies.
    We further classified the non-interacting galaxies into two different morphological classes (Class I and Class III; see text for the details), and we show Class I sources with blue squares and Class III with blue diamonds in this plot.
    The horizontal gray shaded region with a dotted line is the equilibrium value of $\eta_{1500}$ as shown in Figure \ref{fig:eta_dist}.
    The vertical dashed line shows $r_{\rm NN}=1^{\prime\prime}$ which we used as the threshold to discriminate closely paired sources and chance projections in Section \ref{subsubsec:sep_NN}.
    This separation threshold corresponds to $\sim5.8$ physical kpc at $z\sim5.8$.
}
    \label{fig:burstiness}
\end{figure}

\subsubsection{Difference between interacting and non-interacting galaxies}\label{subsubsec:int_vs_noint}
We next investigated the relation between galaxy-galaxy interactions and SFH burstiness.
Figure \ref{fig:burstiness} shows the distribution of sample galaxies in the $r_{\rm NN}$ vs $\eta_{1500}$ plane, with red open squares representing the interacting galaxy sample (Class II + Class II* objects).
Note that red squares whose $r_{\rm NN}$ is smaller than $1^{\prime\prime}$ are Class II galaxies, and we should use only them (Class II but not Class II*) when we want a more conservative estimation of a quantity for interacting galaxies.

To examine the relation between interactions and SFH burstiness, we compared the $\eta_{1500}$ value distributions between interacting and non-interacting galaxy samples, taking measurement uncertainties into account as follows.
We first weighted each sample galaxy by the inverse of its relative $\eta_{1500}$ uncertainty ($\Delta \eta_{1500}/\eta_{1500}$).
The sample galaxies were then divided into two subsamples (interacting or non-interacting), and for each subsample we calculated the weighted count in each $\log(\eta_{1500})$ bin to obtain the uncertainty-weighted $\eta_{1500}$ histogram.
We then normalized each weighted distribution to unity area, and thus obtained the probability distribution functions (PDFs) of the $\eta_{1500}$ parameter for the interacting galaxy sample and the non-interaction sample separately.%

\begin{figure*}
    \includegraphics[width=\textwidth]{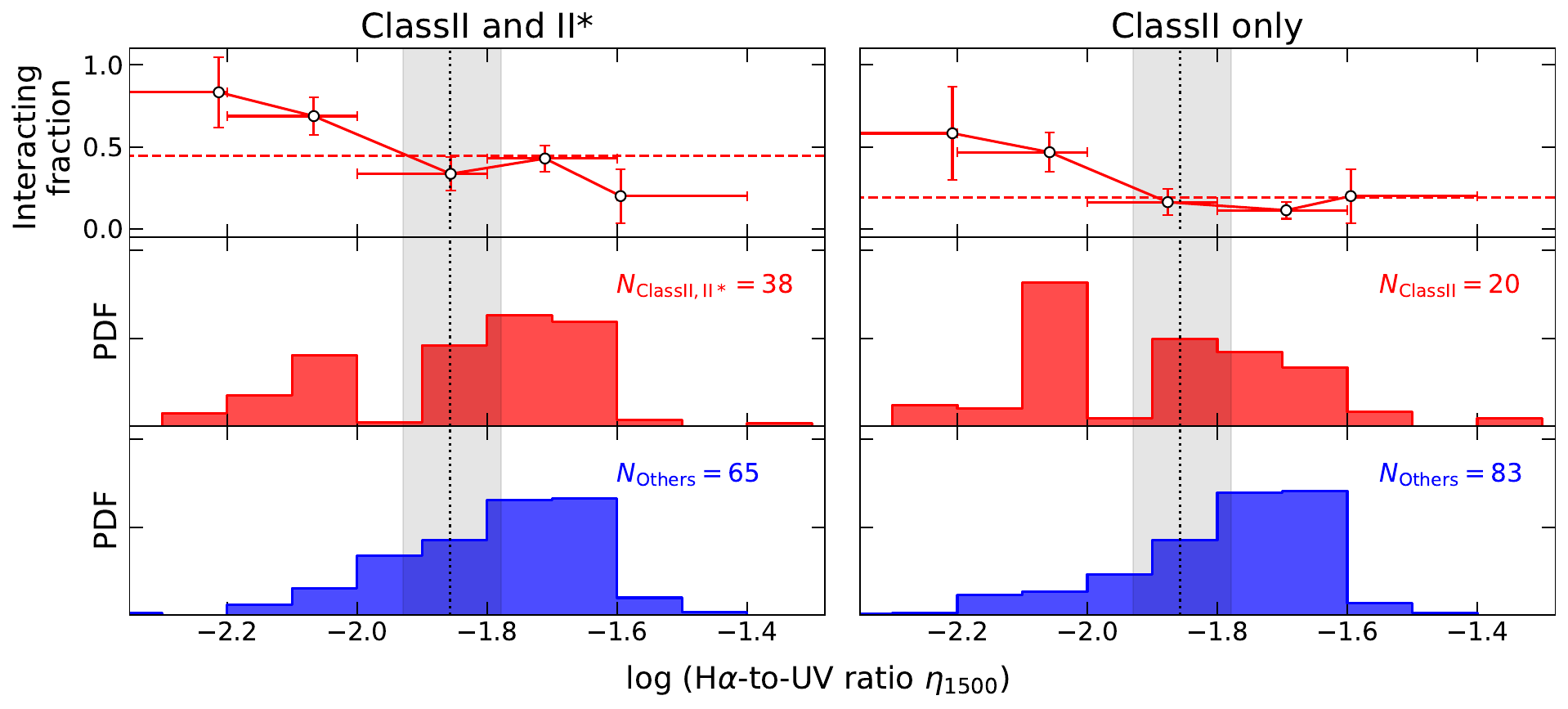}
    \caption{The H$\alpha$-to-UV flux ratio $\eta_{1500}$ PDFs of interacting galaxies (middle: red) and non-interacting galaxies (bottom: blue).
    Shaded region with a vertical dotted line shows the range of equilibrium value of $\eta_{1500}$ as shown in Figure \ref{fig:eta_dist}.
    The top panel shows the fraction occupied by interacting galaxies to the total population against $\eta_{1500}$ values.
    The horizontal dashed line in this panel denotes the overall average of interacting galaxy fraction, thus the $\eta_{1500}$ PDF in interacting galaxies (middle panel) is more biased towards where the fraction is higher than the dashed line.
    Left panel presents these results when the fiducial definition of interacting galaxy sample (ClassII + ClassII*; based on the visual inspection) are used, whilst right panel presents those when more conservative definition of interacting galaxy (only ClassII) is used.
    The results do not change regardless of the definition.
}
    \label{fig:eta_PDF}
\end{figure*}

Figure \ref{fig:eta_PDF} shows the resulting $\eta_{1500}$ PDFs for two definitions of interacting galaxies.
In the left panels, we show the result when we use the fiducial definition of interacting galaxy (i.e., ClassII + II*), while the right panels show the result with the conservative definition (i.e., Class II only).  
The middle and bottom panels in Figure \ref{fig:eta_PDF} show the resulting $\eta_{1500}$ PDFs for the interacting galaxy samples (red) and the non-interacting galaxy samples (blue), respectively.
For brevity, we focus the discussion that follows on results with the fiducial definition (left panels) but note that the more conservative definition (Class II only) in the right panels yields essentially the same conclusions.

Comparing the interacting and non-interacting \bursteta\ PDFs, we note the strong difference at low $\eta_{1500}$ ($\log(\eta_{1500})<-2$) between the interacting and non-interacting populations:  while a significant fraction of interacting galaxies has such low $\eta_{1500}$ values, there are very few non-interacting galaxies in that region.
Indeed, the probability obtained by integrating the PDF over $\log(\eta_{1500})<-2.0$ is $15.7^{+7.3}_{-6.6} \%$ in interacting galaxy sample, whereas that is only $6.8^{+2.7}_{-2.5} \%$ in non-interacting sample \footnote{Uncertainties on the probabilities are estimated by bootstrap resampling.}.
Moreover, the  interacting galaxy sample PDF shows a notable gap at $\log(\eta_{1500})\sim-1.9$, whereas the non-interacting galaxy sample is smoothly distributed in \bursteta. 

To demonstrate the difference between the $\eta_{1500}$ distributions, we also derived the fraction occupied by interacting galaxies to the whole sample in each $\eta_{1500}$ bin, and examined its dependence on $\eta_{1500}$ values.
The top left panel in Figure \ref{fig:eta_PDF} presents the $\eta_{1500}$ dependence of the resulting  interacting galaxy fraction\footnote{The fraction here is not exactly the same as the number fraction of interacting galaxies, because each galaxy is weighted by its $\eta_{1500}$ relative uncertainty.}. It clearly shows that the fraction increases as the $\eta_{1500}$ value decreases, and the fraction is significantly higher than the overall average fraction (horizontal dashed line in the top panel) in the lowest-$\eta_{1500}$ region, and reaches $\gtrsim50 \%$ in the lowest-$\eta_{1500}$ population.
The vertical error bars in this plot are estimated based on Poisson statistics.
This plot indicates that the $\eta_{1500}$ distribution for interacting galaxies has a different shape than non-interacting galaxy sample, and particularly so in the lowest-$\eta_{1500}$ region. 
The interacting galaxies have $\eta_{1500}$ distribution that is more biased towards the lowest-$\eta_{1500}$ region.

These results do not change regardless of the definition of "interacting galaxy sample": the same trends can be seen regardless of whether we use the canonical Class II + Class II* (left panels)  interacting galaxy sample, or the more conservative Class II-only definition (right panels). 

These results suggest that the local environment of galaxies, or galaxy-galaxy interactions, indeed enhances  SFH burstiness in high-$z$ low-mass galaxies. In particular, it enables faster quenching.
Given that the H$\alpha$-to-UV ratio $\eta_{1500}$ is identical to the ratio of short-timescale SFR over the long-timescale SFR, the larger fraction of lowest-$\eta_{1500}$ population in interacting galaxy sample should suggest interacting galaxies typically experience shorter timescale quenching. 
The lowest $\eta_{1500}$ values such as $\log(\eta_{1500})\lesssim -2.1$ typically require quenching timescales of $\tau\lesssim100$ Myr (see Appendix \ref{apx:toy_model}).
Moreover, the gap seen in the interacting galaxy PDF also indicates more rapid transition from the high-$\eta_{1500}$ state (i.e., rising SFR phase) to low-$\eta_{1500}$ state (i.e., declining SFR phase), because the small number of intermediate (or equilibrium) $\eta_{1500}$ population should suggest that galaxies spend much shorter time in the transition phase.
On the other hand, the presence of the high-$\eta_{1500}$ peak in non-interacting galaxy sample (blue) can suggest galaxies without interactions also may experience a rapid increase of star formation activities.
Although, combined with the absence of lowest-$\eta_{1500}$ population and the continuous PDF distribution, non-interacting galaxies should be less likely to experience rapid SF declining and their SFHs should be less bursty.  We explore this interpretation further in the next Section  using toy models.

\subsubsection{Interpretation with toy models}\label{subsubsec:toymodels}
\begin{figure}
	\includegraphics[width=\columnwidth]{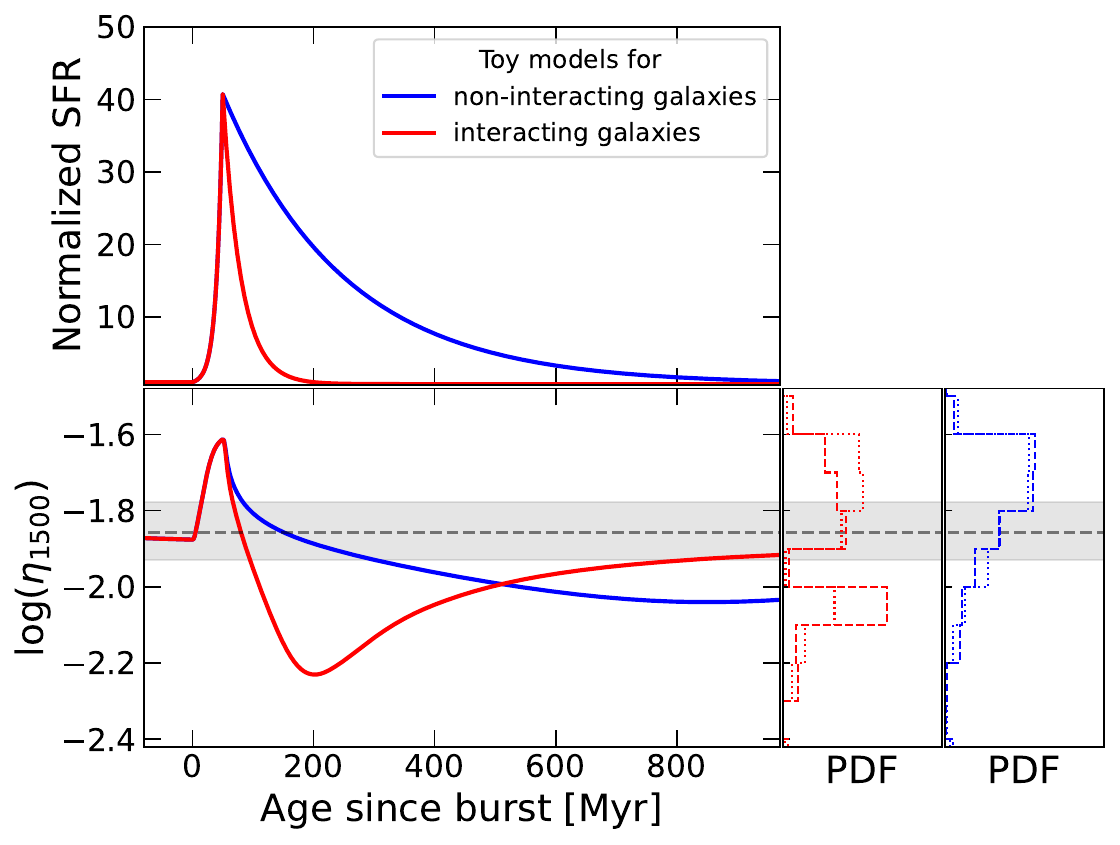}
    \caption{Toy models of $\eta_{1500}$ evolution to describe the difference between the  $\eta_{1500}$ distributions of the interacting and non-interacting galaxy samples.
    We assumed two different SFHs (top panel) and calculated the corresponding $\eta_{1500}$ histories  (bottom left) that can explain the observed  $\eta_{1500}$ PDFs (bottom right -- taken from  Figure \ref{fig:eta_PDF}).
    Both scenarios are characterized by a rapid SFR rise but they differ in how star formation declines:  the $\eta_{1500}$ PDF for interacting galaxies (red histograms in bottom right) can be explained with a burst that undergoes a  rapid decline (red curve, top panel). In contrast, that of non-interacting galaxies (blue histograms in bottom right) can be modeled by a burst followed by a slow, gradual decline (red, top).
    Note that we do not correct for sample incompleteness and thus the $\eta_{1500}$ evolution of each model in the bottom left panel is not expected to exactly match the observed PDFs in the bottom right panel (see Section \ref{subsec:sample_bias} for the details).
    The SFR values in the top panel are normalized by the underlying constant SFR level (see Appendix \ref{apx:toy_model} for the details).
}
    \label{fig:eta_FSPS}
\end{figure}

To visualize this interpretation of $\eta_{1500}$ PDF differences, we built toy models of $\eta_{1500}$ evolution under two different SFH assumptions.
Figure \ref{fig:eta_FSPS} shows the two toy models to describe this difference between the $\eta_{1500}$ PDFs of interacting galaxy sample (red) and non-interacting sample (blue).
The $\eta_{1500}$ PDF of interacting galaxies (bottom right, red) can be explained by a bursty SFH that has a rapid rise and a subsequent fast quenching episode (top left, red).
If we assume a SFH characterized by a rapid rise and fast quenching, the $\eta_{1500}$ value increases to $\log\eta_{1500}\sim-1.6$ as the SFR is increasing but $\eta_{1500}$ decreases down to $\log\eta_{1500}\sim-2.2$ rapidly after the SFR starts to decrease (bottom left, red).
Since this transition occurs quickly, the probability of finding galaxies in the phase that has intermediate $\eta_{1500}$ values ($\log\eta_{1500}\sim-1.9$) would be small.
After the fast transition, the $\eta_{1500}$ value stays in the low value for a few 100 Myrs, which enlarges the probability of finding this lowest-$\eta_{1500}$ population, and there emerges another peak at this low $\eta_{1500}$ value, with a gap between the two peaks.
The timescale of the rising/declining phase is quite short (in this model we assumed $\tau=10$ Myr and $\tau=30$ Myr for the rising and declining phase, respectively), and a galaxy can experience this type of burst several times within the age of Universe at $z\sim5.8$ ($\sim 1$ Gyr).
Note that the timescale for residing  in the high-$\eta_{1500}$ region is also short, but these galaxies have high SFR values and are easy to detect, which  makes the peak at high $\eta_{1500}$ value distinctive.

On the contrary, the $\eta_{1500}$ PDF of non-interacting galaxies (bottom right, blue) can be explained by a SFH with a rapidly rising SFR event that is then followed by a slow, gradually declining {\it cooldown} phase (top left, blue).  In this model, the $\eta_{1500}$ value also increases to $\log\eta_{1500}\sim-1.6$ at first but then decreases slowly and only to as low as  $\log\eta_{1500}\sim-2.0$ (bottom left, blue).
In this scenario, the transition from the high-$\eta_{1500}$ to low-$\eta_{1500}$ value is only gradual, and this ensures that -- in contrast to the fast-quenching scenario -- there is no gap here in the $\eta_{1500}$ PDF.
Due to the long timescale of the declining phase (in this model we assumed $\tau=300$ Myr), it is hardly possible for a galaxy at this redshift to experience this episode several times, and therefore this behaviour cannot be an episodic burst.
Detailed configurations of our toy models and further detailed discussion are given in Appendix \ref{apx:toy_model}.

In summary, the observations and our toy models suggest that the difference between interacting and non-interacting galaxy  $\eta_{1500}$ PDFs points to a difference in the typical SFHs in these two populations.
Interacting galaxies could typically experience bursty SFHs, characterized by a rapid rise of SFR and the subsequent fast quenching.
Particularly, as compared to non-interacting galaxies, the timescale of the quenching phase is expected to be shorter in the burst episode that interacting galaxies experience.
Our toy model suggests that in interacting galaxies both the rising-SFR and quenching phases have timescales of a few tens of Myr and shorter than 100 Myrs. 
On the other hand, non-interacting galaxies can also have high-$\eta_{1500}$ values, which suggests a rapidly rising-SFR phase similar to that in the interacting galaxies. However, in these isolated galaxies the mechanism responsible cannot be galaxy-galaxy interactions, suggesting the presence of another physical mechanisms responsible for triggering the rapidly rising SFR phase.
Moreover, non-interacting galaxies are not quenched as rapidly as interacting galaxies, with slow quenching timescales of $>$100 Myr (see also Appendix \ref{apx:toy_model}).
Given this long quenching phase, it is unlikely that a typical galaxy experiences such a star formation episode several times within the age of Universe ($\sim$1~Gyr at the redshift of our sample).  Therefore, the SFHs of non-interacting galaxies cannot be seen as bursty, in marked contrast to the rapid bursts present in the  SFHs of interacting galaxies.

\subsection{High interacting-galaxy fraction}\label{subsec:merger_frac}
In this work, we consider galaxies with a high-$z$ companion whose source-plane separation is $<1$ arcsec to be interacting galaxies.
Using the number of these galaxies, we can estimate the fraction of interacting galaxies in this sample.
Table \ref{tab:merger_frac} shows the number of galaxies in each morphological class determined in Section \ref{subsec:morph_cls}.
When we use the fiducial sample of interacting galaxies (Class~II + Class~II*) to calculate the interacting galaxy fraction ($f_m$), the fraction is $f_m=38^{+4}_{-5}$ \%.
The uncertainty on this fraction is estimated by bootstrap resampling.
This fraction is higher than typical merger fractions of low-mass galaxies at low-$z$ \citep[$\lesssim10$ \%; e.g.,][]{Conselice2008,Casteels2014,Robotham2014,Besla2018}, which  indicates that merging of galaxeis can be important for the mass growth in high-$z$ low-mass galaxies \citep[see also, e.g.,][]{Asada2023,Witten2023}.
Combined with the results in Section \ref{subsec:burstiness_and_interaction}, the high merger fraction or frequent galaxy-galaxy interaction events could cause the fluctuating star formation activities in high-$z$ low-mass galaxies in general.

It is hard to tell apart robustly between interacting galaxies and multiple stellar components within a single larger galaxy without kinematic information, and some of the closely paired objects in our sample could be star forming clumps in individual galaxies.
Nevertheless, the separations between these paired galaxies are typically $\sim0.3$ arcsec (Figure \ref{fig:burstiness} top), which corresponds to $\sim2$ physical kpc at $z=5.8$, and our sample galaxies are mainly low-mass galaxies with $M_\star \sim 10^8\ M_\odot$ (Figure \ref{fig:sSFR_mstar}). Thus, such closely paired objects are more likely to be interacting or merging galaxies rather than clumps within larger objects.

\begin{table*}
    \centering
    \caption{Interacting galaxy fractions in this work. 
    Numbers of sources for each morphological classes based on visual inspections in Section \ref{subsubsec:VI} are shown.
    The fiducial interacting galaxy fraction $f_m$ is calculated assuming Class II and II* sources are interacting galaxies.
    The most conservative fraction $f_m^{\rm cons}$ is derived using only Class II galaxies as the interacting sample, whereas the most aggressive fraction $f_m^{\rm agg}$ is using all the Class II, II*, and III galaxies (see the text for details). Uncertainties on the interacting galaxy fractions are estimated by bootstrap resampling.
  	}
    \label{tab:merger_frac}
    \begin{tabular}{lcccccccc} 
		\hline\hline
		Field & \# of sources & Class I & Class II & Class II* & Class III & $f_m$ & $f_m^{\rm cons}$ & $f_m^{\rm agg}$\\
		\hline
		CLU  & 47 &  16 & 17 & 6 & 8 & $49^{+9}_{-6}$ \% & $36^{+6}_{-9}$ \% & $66^{+6}_{-6}$ \% \rule[-2mm]{0mm}{4mm} \\
		NCF & 76 & 44 & 6 & 18 & 8 & $32^{+5}_{-5}$ \% & $8^{+4}_{-3}$ \% & $42^{+5}_{-5}$ \% \rule[-2mm]{0mm}{4mm} \\
        \hline
		All & 123 & 60 & 23 & 24 & 16 & $38^{+4}_{-5}$ \% & $19^{+3}_{-4}$ \% & $51^{+5}_{-5}$ \% \rule[-2mm]{0mm}{4mm} \\
		\hline
	\end{tabular}
\end{table*}

Even if we use the most conservative definition of interacting galaxy (i.e., only Class II objects) to calculate the merger fraction, the fraction is still high, $f_m^{\rm cons} = 19^{+3}_{-4}$ \%.
As discussed in Section \ref{subsec:morph_cls}, using this definition of interacting galaxy should underestimate the number of closely paired galaxies.
Particularly in the NCF field, six apparently paired sources (1200422, 1200555, 1200762, 1202210, 1204406, and 1204730) are blended and the source detection failed to segment them into several sources.
These sources are classified as Class II*, and are not counted as mergers in this most conservative calculation.
Therefore, the conservative merger fraction $f_m^{\rm cons}$ should be regarded as the lower limit, yet the value is still higher than expected from previous studies in the low-$z$ universe.

Moreover, Class III sources in this work have been regarded as non-interacting galaxies, but these galaxies show complex morphologies that may be indicative of galaxy-galaxy interactions.
Regarding all the Class II, Class II*, and Class III sources as interacting galaxies should thus give the most aggressive estimate of the merger fraction, that can be regarded as the upper limit on this quantity.
The resulting aggressive estimation of the merger fraction, ($f_m^{\rm agg}$) is $51^{+5}_{-5}$ \%.
Thus, up to about a half of the observed low-mass galaxies at $z\sim5.8$ may be experiencing a galaxy-galaxy interaction event in progress.

The merger fraction here depends only on the projected separation in the sky, and thus there could be contamination by chance projections from galaxies at significantly different line-of-sight distance.
However, we required  separation smaller than 1 arcsec, and this small threshold should minimize the effect of contamination.
We also statistically evaluated the effect of the chance projection contamination as follows.
We randomly distributed the number of sample galaxies detected in NCF (76) field over the surveyed sky area of NCF field, and measured the separations from the nearest neighbor for each randomly positioned sources.
We then calculated the number of sources having a closely projected companion within 1 arcsec.
We repeated this process 3000 times to calculate the average number of these closely placed sources, and found that the average is only $0.87$.
Therefore, the expected number of finding close ($<1$ arcsec) chance projections  is less than 1 in this sample. Consequently, close pairs with a separation of less than 1 arcsec should be regarded as physically interacting systems. This conclusion is consistent with the presence of a  gap at $\sim$ 1 arcsec in the $r_{\rm NN}$ distribution (top panel in Figure \ref{fig:burstiness}), which suggests that nearest neighbors whose separation is larger than 1 arcsec are mostly chance projection but those smaller than 1 arcsec should be physically interacting systems.

\begin{figure}
	\includegraphics[width=\columnwidth]{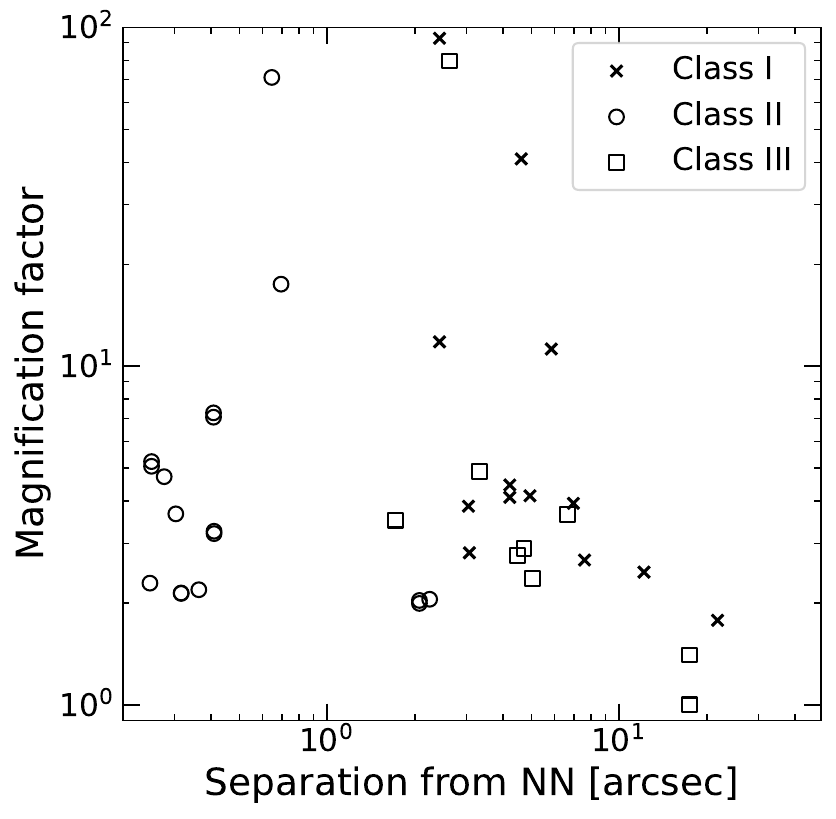}
    \caption{Magnification factor vs. separation from the nearest neighbor ($r_{\rm NN}$) plot.
    Only sample galaxies in the CLU field are shown.
    Crosses, circles, and squares show the morphological classes of Class I, Class II or II*, and Class III, respectively.
}
    \label{fig:magnif_sep}
\end{figure}

Another possible concern is lensing bias.
Since galaxies in the CLU field are magnified by the gravitational lensing effect, the high fraction of interacting galaxies may be because highly magnified galaxies are spatially resolved and their internal structures are confused for multiple sources.
If this were the case, Class II or Class II* sources would be biased to highly magnified sources. However, this is not the case: 
In Figure \ref{fig:magnif_sep}, we show the source distribution in the $r_{\rm NN}$ vs. magnification factor plane.
As seen from this figure, Class II or Class II* sources do not necessarily have high magnification factors, and highly magnified sources are not always classified as interacting galaxies.
This indicates that the gravitational lensing effect does not bias the merger fraction calculation.

\section{Discussion}\label{sec:discussion}

\subsection{Galaxy-galaxy interactions and their effects on SFHs}

Our analysis has found that a large fraction of low-mass galaxies at $4.7\lesssim z\lesssim6.5$
have \bursteta values that are out of equilibrium, consistent with rapidly changing — i.e., bursty —  star formation histories. Furthermore, many ($\sim$ 40 percent) of our galaxies are found in close association ($<6$ kpc in the source plane) with each other.  Because chance projections can be expected to be very rare ($<2$ percent of cases), these closely-associated galaxies are very likely in interacting pairs or groups.  

Whether these pairs or groups will merge with each other remains an open question since we do not yet know their DM halo masses or kinematics.  However,  given their physical proximity, their members are likely to be physically interacting — an assertion that is supported by the fact that the star formation histories we infer from the distributions of \bursteta\ values are different for the interacting and non-interacting populations.  Specifically, while both populations have rising phases that rise rapidly (timescales of a few tens of Myr), they differ in their subsequent evolution: in isolated galaxies, the post-peak star formation rates decline slowly (timescales of $>100$ Myr).  In contrast, in interacting galaxies the quenching of star formation is much more rapid (timescales of tens of Myr) as evidenced by the presence of very low \Ha-to-UV ratios of $\log(\eta_{1500})<-2$ that are found particularly in this  population.

We can speculate that the rapid rise in SFR experienced by the interacting galaxies may be caused by their physical encounters.  This would be consistent with the elevated star formation rates seen in low-$z$ interacting galaxies \citep[e.g.,][]{Scudder2012,Patton2013,Barrera-Ballesteros2015}, although the bursts of star formation in our high-$z$ low-mass galaxies appear to  more rapid than what is seen at low redshifts \citep[e.g.,][]{Ellison2022}.  We cannot tell from our present analysis what causes the very rapid quenching of star formation in the interacting galaxies. What is clear, however, is that the quenching mechanism in our interacting galaxies must be different from the mechanism responsible for the more gradual SFR decline experienced by the isolated galaxies.  If our Class I galaxies are truly isolated (rather than paired with undetected companions that are below our luminosity limit), then the rapid rise of star formation we observe in them are not due to galaxy-galaxy interactions and must be triggered by another, separate type of triggering mechanism such accretion of gas clumps from outside the galaxy or violent disk instabilities \citep{Agertz2009, Dekel2009}.

\subsection{Merger fractions across the cosmic time}
As mentioned earlier, without information about galaxy-pair kinematics and total (DM halo) masses we cannot yet conclude that our paired galaxies will become galaxy-galaxy mergers.  However, assuming that they do merge, we can examine some implications.  Taking the $\sim$40\% pair fraction to represent the merger fraction, we note that the merger fraction in this work is significantly higher than that of similarly low-mass galaxies in the low-$z$ Universe \citep[$\lesssim10\ \%$; e.g.,][]{Robotham2014}, and this indicates a cosmic evolution of low-mass galaxy merger fraction.
Indeed, some simulations also have shown a different merger fraction evolution between high-mass and low-mass galaxies, and the merger fraction among low-mass galaxies can be higher at high redshifts than in the local Universe \citep[e.g.,][]{Husko2022,Chamberlain2023}.
In the low-$z$ Universe, low-mass galaxies are expected to mainly experience extremely high-mass ratio mergers, with $m_2/m_1>10$ \citep[e.g.,][]{O'Leary2021}.
This is probably because of the difficulty in robustly identifying low-$z$ low-mass galaxies that are free to merge with each other rather than being immersed in the halos of more massive systems. 
When in such massive systems, the low-z low-mass galaxies can be expected to accrete into the outskirts of their much more massive central galaxies \citep[see, e.g., ][;Williams et al. in prep]{Huang2018}. 

In contrast to the situation at low redshifts, most of the galaxies in our interacting population are in systems where their companions have similar masses: indeed, among the 23 Class II galaxies in our sample, 20 galaxies have a companion whose stellar mass ratio is less than 10.
They (or their companions) could therefore be the most massive galaxies in their DM halos.
If these pairs and groups are embedded in such common DM halos then they can be expected to sink to the halo centres and merge rapidly as dynamical friction acts most strongly on the highest-mass objects \citep[see, e.g.,][for examples of this at lower redshifts and higher masses]{Jones2000,Sawicki2020}.
Therefore, the galaxy-galaxy interaction phenomenon among low-mass galaxies in the high-$z$ Universe could be qualitatively different from that at low-$z$, and the significant evolution of merger fraction across cosmic time could be attributed to a fundamental difference in the nature of the mergers.

\subsection{Effect of selection bias and sample incompleteness}\label{subsec:sample_bias}
This work used three different sample selection criteria to select galaxies at $4.7\lesssim z\lesssim6.5$.
As discussed in Section \ref{sec:sample}, we tried to mitigate the bias in $\eta_{1500}$ of our sample by avoiding phot-$z$ uncertainty cut, though some selection bias or sample incompleteness may remain.
We did not correct for the incompleteness in this work, but the incompleteness is not expected to affect our results qualitatively.

One of the key results of this work is the difference of $\eta_{1500}$ PDFs between interacting and non-interacting galaxies.  However,  sample incompleteness should affect  both the interacting and non-interacting galaxy samples similarly, thus  selection bias cannot explain the difference in the $\eta_{1500}$ PDFs between these populations.

Qualitatively, galaxies at the peak of SF bursts would be bright and easy to detect whereas those in the declining phase would be difficult to detect, thus  selection bias should affect the shape of $\eta_{1500}$ PDF.
Figure \ref{fig:eta_PDF} shows that the PDF peak in the high-$\eta_{1500}$ region is higher than that in the low-$\eta_{1500}$ region, but this difference of PDF peak height is expected to be affected by the selection bias.
For example, in the toy model for non-interacting galaxies shown in Figure \ref{fig:eta_PDF} (blue curves), galaxies spend more time in the low-$\eta_{1500}$ region and one might expect the corresponding PDF to be dominated by this low-$\eta_{1500}$ population, in apparent conflict with the observed PDF (blue PDFs).
However, galaxies in the low-$\eta_{1500}$ region in this model are aged a few 100 Myrs after the peak of the burst, with SFRs already considerably lower than at the peak of the burst (see top panel of Fig.~\ref{fig:eta_PDF}). They are thus expected to be faint both in rest-frame UV and Ha and difficult to detect.
This affects the observed shape of the PDF, suppressing the low-$\eta_{1500}$ region and biasing the PDF towards the high-$\eta_{1500}$ population.
On the other hand, in the toy model for interacting galaxies (red curves), galaxies transit from high- to low-$\eta_{1500}$ region so quickly that even low-$\eta_{1500}$ galaxies can be still bright in rest-frame UV, which makes them detectable via the LBG selection, and therefore the low-$\eta_{1500}$ population is expected to be more visible in the fast quenching model.
Correcting for such sample incompleteness is thus required to quantitatively determine the SFH parameters (e.g., rising/declining timescale $\tau$) by fitting the $\eta_{1500}$ distribution to the observed one \citep[see e.g.,][]{Emami2019}, but that also requires a larger sample of galaxies and we leave it for a future work where the full data set from CANUCS program is available.

\section{Conclusions}
\label{sec:conclusions}

The star formation activities in low-mass galaxy population in the  high-$z$ Universe are one of the keys to reveal the cosmological evolution of galaxies.
The galaxy population has been expected to typically experience fluctuating SFHs (bursty SFHs) rather than smooth and steady SFHs by simulations and observations at lower-$z$ \citep[e.g.,][]{Sparre2017,Emami2019}.
Early results from JWST observations have identified some evidence of such fluctuating star formation activities in high-$z$ low-mass galaxies including extremely high equivalent width emission lines or prominent Balmer breaks at $z\gtrsim4$ \citep[e.g.,][]{Strait2023,Withers2023}.
However, the physical origin of the fluctuation is still unknown, and a statistical study has been lacking until now.

In this paper, we used deep and high spatial resolution JWST/NIRCam imaging data to, for the first time,  investigate the SFH burstiness and the local environments of low-mass galaxies at $z\sim5-6$.
We exploited NIRCam observations in a gravitational lensing cluster field, MACS0417, taken as a part of CANUCS program.
We obtained a sample of 123 galaxies at $z\sim4.7-6.5$ based on their colours, and examined their star formation burstiness using the H$\alpha$-to-UV flux ratio ($\eta_{1500}$).
We also visually selected interacting galaxies, and probed the relation between star formation burstiness and galaxy-galaxy interactions.
Our main results are:
\renewcommand{\theenumi}{\arabic{enumi}}
\begin{enumerate}
    \item The $\eta_{1500}$ (i.e., burstiness) values of our samaple galaxies extend from below to above the range that steady, smooth SFHs can achieve.
    60\% of the sample galaxies have $\eta_{1500}$ that deviates from the equilibrium range at $>1\sigma$ level, suggesting that the majority of high-$z$ low-mass galaxies in the sample experience fluctuating SFHs (Section \ref{subsec:burstiness_and_interaction} and Figure \ref{fig:eta_dist}). This fraction may well be even higher if we accept that some of our galaxies have insufficient S/N to meet our $>1\sigma$ criterion or that some bursty galaxies at the transition phase will show equilibrium-like \bursteta values.
    \item Comparing the $\eta_{1500}$ distributions of the interacting galaxy sample and the non-interacting galaxy sample, interacting galaxies are more likely to have particularly low $\eta_{1500}$ values ($\log(\eta_{1500})<-2.0$) as compared to non-interacting galaxies (Figure \ref{fig:eta_PDF}).
    This difference can be explained by a difference in the typical SFHs among the two populations.
    Although both populations seem to experience a rapid upturn of star formation activity, interacting galaxies could typically experience faster quenching while non-interacting galaxies only gradually decrease their star formation activities (Section \ref{subsec:burstiness_and_interaction}). This suggests that galaxy-galaxy interactions enhance star formation burstiness in the galaxy population.
    \item The interacting galaxy number fraction in the sample is higher as compared to e.g., typical merger fraction of star-forming galaxies in the  low-$z$ Universe ($\lesssim10$ \%).
    $38^{+4}_{-5} \%$ of sample galaxies appear to have a similar-redshift close companion within 1 arcsec, which corresponds to $\sim5.8$ kpc at $z\sim5.8$ (Table \ref{tab:merger_frac}).
    Given the effect of galaxy-galaxy interaction on SFH burstiness, the high interaction fraction could cause the bursty SFHs in low-mass, high-$z$ galaxies to be a nearly ubiquitous phenomenon (Section \ref{subsec:merger_frac}).
\end{enumerate}

Our results have shown the significant role of bursty star formation in low-mass high-$z$ galaxies and given an insight into the physical origin of the SFR fluctuations.
In low-mass galaxies in the high-$z$ Universe, galaxy-galaxy interactions could be one of the leading causes of the bursty SFHs.  Our results show that this is also be the case at high redshift. 
Further quantitative analyses require a larger sample of this galaxy population, which will be available with our full CANUCS data set.
Furthermore, spectroscopic follow-up campaigns will reveal their properties unambiguously, and give a more fundamental view of how galaxy-galaxy interactions affect the cosmological evolution of low-mass galaxies.

\section*{Acknowledgements}


We thank the anonymous reviewer for the comments and suggestions to improve this paper.
This research was enabled by grant 18JWST-GTO1 from the Canadian Space Agency, and funding from the Natural Sciences and Engineering Research Council of Canada.
YA is supported by a Research Fellowship for Young Scientists from the Japan Society of the Promotion of Science (JSPS).
MB acknowledges support from the ERC Grant FIRSTLIGHT, Slovenian national research agency ARRS through grants N1-0238 and P1-0188, and the program HST-GO-16667, provided through a grant from the STScI under NASA contract NAS5-26555.

This research used the Canadian Advanced Network For Astronomy Research (CANFAR) operated in partnership by the Canadian Astronomy Data Centre and The Digital Research Alliance of Canada with support from the National Research Council of Canada the Canadian Space Agency, CANARIE and the Canadian Foundation for Innovation.

\section*{Data Availability}


Raw JWST data used in this work will be available from the {\it Mikulski Archive for
Space Telescopes} (\url{https://archive.stsci.edu}), 
at the end of the 1-year proprietary time (doi: 10.17909/ph4n-6n76). Processed data products will be available on a similar timescale at \url{http://canucs-jwst.com}.



\bibliographystyle{mnras}
\bibliography{reference} 



\newpage

\appendix

\section{Postage stamps of the sample galaxies}\label{apx:postage_stamps_all}
\begin{figure*}
 \centering
 \includegraphics[width=1\textwidth]{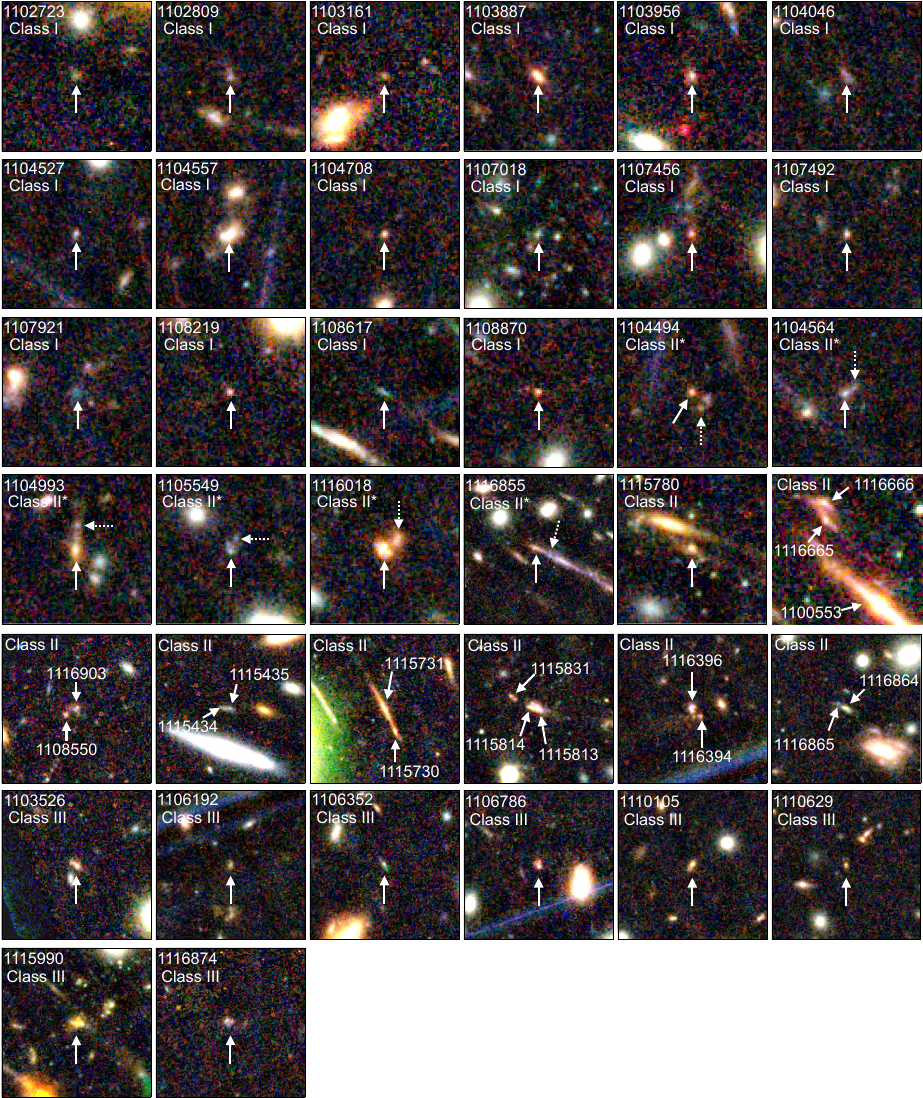}
 \caption{RGB cutout images of CLU sample galaxies. White solid arrows point the sample galaxy in this work.
 The cutouts are 3 arcsec on the side, and we used all the NIRCam filters in R, G, B colors (red: F356W+F410M+F444W, green: F200W+F277W+F356W, blue: F090W+F115W+F150W).
 White dashed arrows in panels for Class II* sources point the close pair companions of the sample galaxy (see Section \ref{subsubsec:VI} for the details).
 }
 \label{fig:postage_stamps_all_CLU}
\end{figure*}

In this appendix, we show the RGB composite images of all the 123 sample galaxies in this work.
We also show the morphological classification based on our visual inspection in Section \ref{subsubsec:VI}.
Figure \ref{fig:postage_stamps_all_CLU} shows the 47 galaxies in CLU field, and Figure \ref{fig:postage_stamps_all_NCF1} shows 76 galaxies in NCF field.
The cutouts are 3 arcsec on the side, and white arrows point the sample galaxies in this work.
White dashed arrows in panels of Class II* sources indicate the close pair companion of the sample galaxy (see Section \ref{subsubsec:VI} for the detail).

\begin{figure*}
 \centering
 \includegraphics[width=1\textwidth]{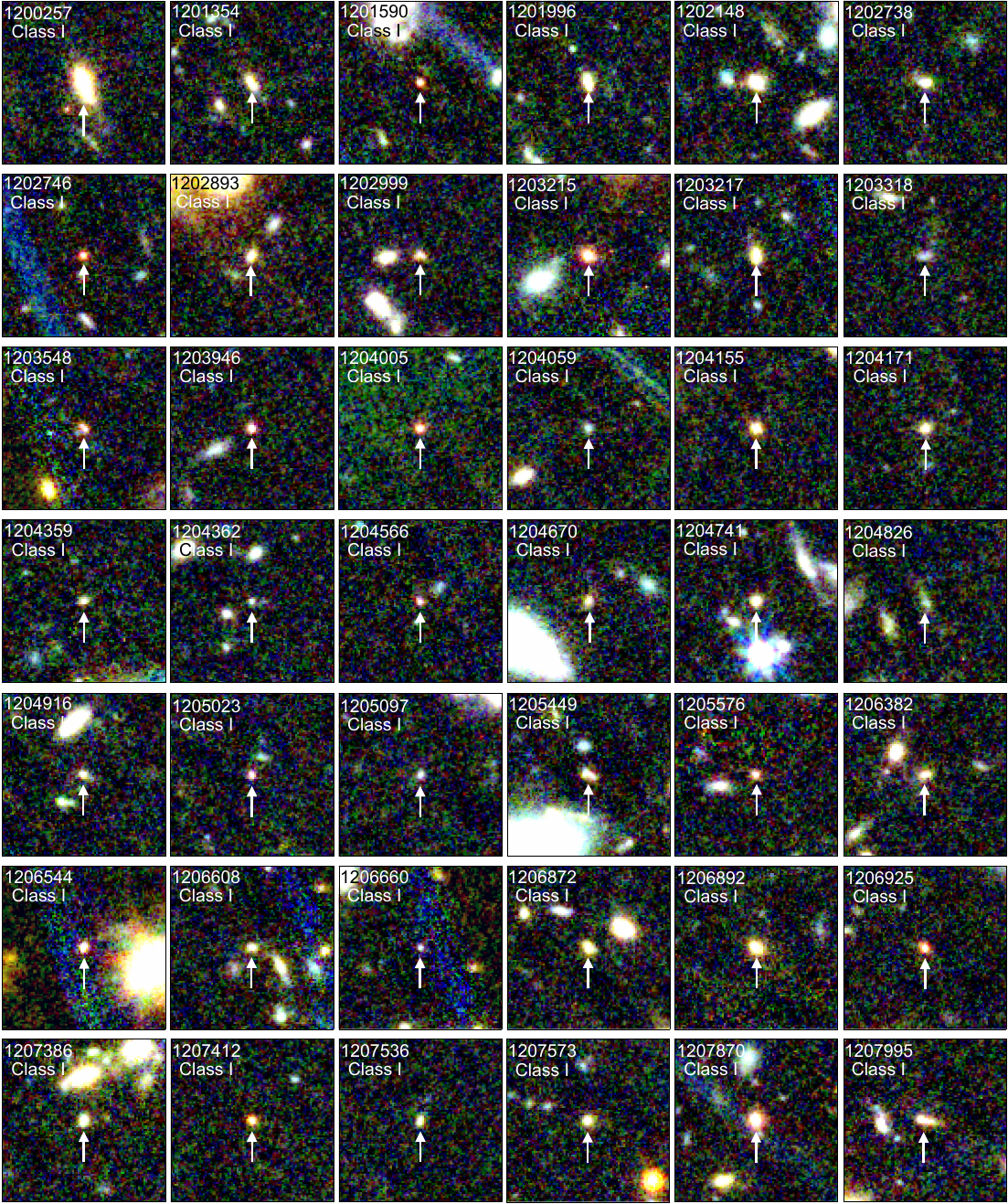}
 \caption{Same as Figure \ref{fig:postage_stamps_all_CLU} but for NCF sample galaxies.}
 \label{fig:postage_stamps_all_NCF1}
\end{figure*}

\begin{figure*}
 \centering
 \includegraphics[width=1\textwidth]{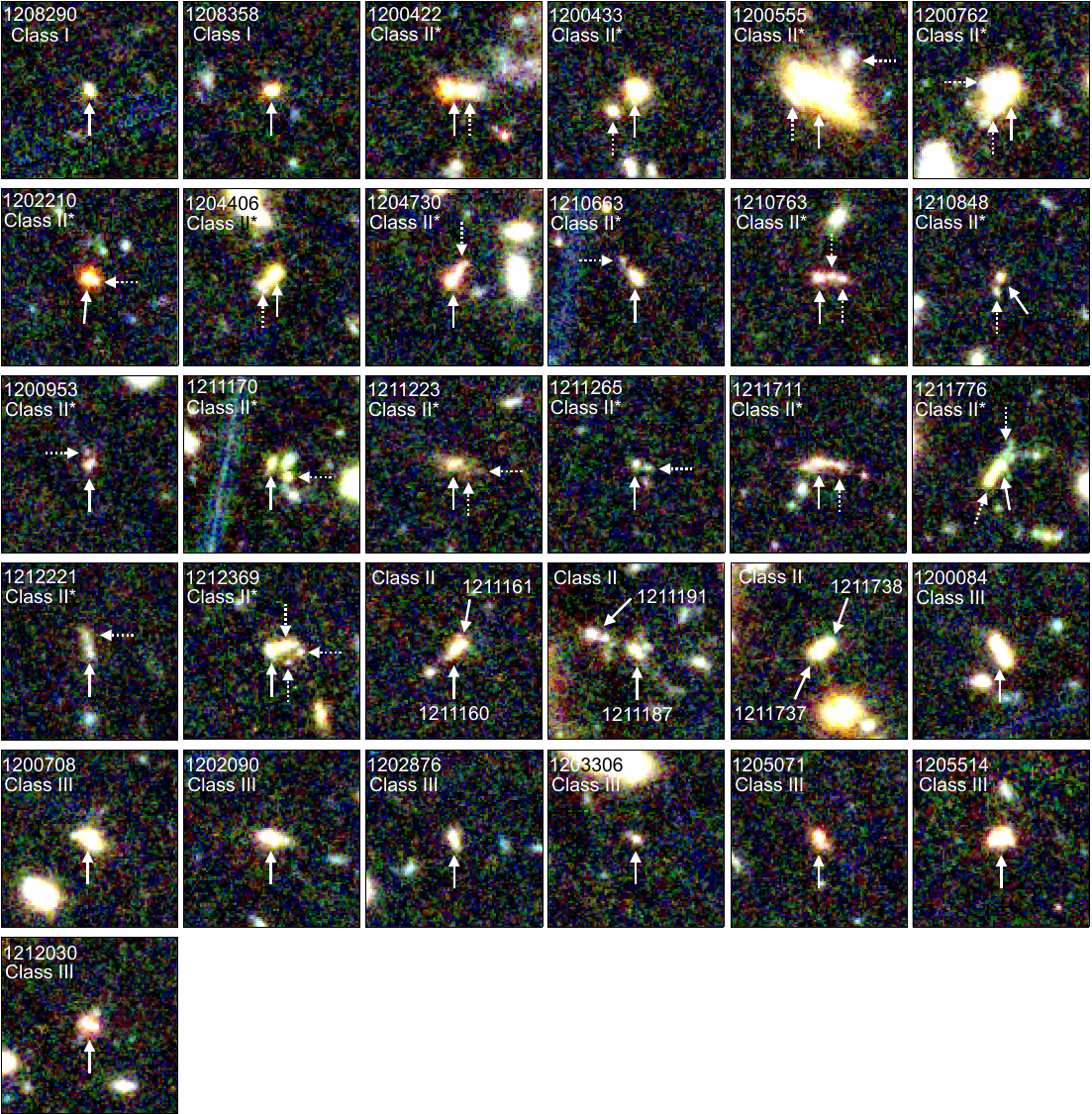}
 \contcaption{}
 \label{fig:postage_stamps_all_NCF2}
\end{figure*}

\section{Toy models of $\eta_{1500}$ evolution}\label{apx:toy_model}
In Section \ref{subsec:burstiness_and_interaction} and Figure \ref{fig:eta_FSPS}, we discussed the time evolution of $\eta_{1500}$ value under different SFH assumptions using two toy models.
In this Appendix, we give the detailed configurations of the two models.

The (dust-corrected) H$\alpha$-to-UV ratio has been suggested as a proxy of the burstiness of the SFHs in a galaxy population \citep[e.g.,][]{faisst_recent_2019}.
If the SFH is smooth and steady for a long ($\gtrsim100$ Myr) timescale, the H$\alpha$-to-UV ratio converges to a equilibrium value, but the equilibrium value can vary depending on other model assumptions such as stellar metallicity or IMF.
\citet{Mehta2022} exploited various set of assumptions to examine the possible range of this equilibrium value, and found $\eta$ should be $1/85<\eta<1/60$ for a steady SFH.
The driving factor of the equilibrium $\eta$ variance is mettalicity, and they considered a metallicity range of $\log(Z/Z_\odot)=-2$ to 0.
In this work, we assumed the metallicity of $\log(Z/Z_\odot)=-1$, which is consistent with the recent metallicity measurements with JWST/NIRSpec of similarly high-$z$ low-mass galaxies \citep[e.g.,][]{Curti2023,Nakajima2023}.

We used the Flexible Stellar Population Synthesis library \citep[FSPS;][]{Conroy2009} to generate the galaxy spectrum templates for a given set of assumptions.
We fixed the stellar and gas-phase metallicity to $\log(Z/Z_\odot)=-1$, and adopted the consistent IMF \citep{Chabrier2003}, and no dust attenuation is applied in the model spectrum.
To model a burst of SF, we prameterized the SFH in a similar way as \citet{Emami2019}, which modeled the rise and decline of a SF burst as an exponential in time, motivated from some hydrodynamical simulations \citep[e.g.,][]{Hopkins2018}.
Namely, we parameterized the SFH as
\begin{eqnarray}\label{eqn:asymmetric_exp}
    \textrm{SFR}(t) = 
    \left\{
    \begin{array}{ll}
    C & (t \leq t_0)\rule[-3mm]{0mm}{6mm},\\
    A \left(\exp{\left[\frac{t-t_0}{\tau_1}\right]} -1 \right) + C & (t_0 < x \leq t_0 + D_1)\rule[-3mm]{0mm}{6mm},\\
    A \left( \exp{\left[-\frac{t-(t_0+D_1+D_2)}{\tau_2}\right]} -1 \right) + C & (t_0 + D_1 < x)\rule[-3mm]{0mm}{6mm},\\
    \end{array}
    \right.
\end{eqnarray}
where $C$ is the underlying constant SF, $\tau_1$ ($\tau_2$) is the $e$-folding time of rising (declining) SF, $D_1$ ($D_2$) is the duration of rising (declining) phase, $t_0$ is the age when the burst starts, and $A$ is the normalization factor of the burst amplitude.
This parameterization is the same as that by \citet{Emami2019}, but allowing to have different $e$-folding time in rising and declining phase.
For simplicity, we fixed the duration as $D_i=5\tau_i$ ($i=1,2$), and introduced an exponential burst on the top of a underlying constant star formation activity aged 500 Myr ($t_0=500$ Myr) assuming the peak of the burst has $\sim40$ times higher SFR value than the underlying constant SFR (i.e., $A=40Ce^{-5}$).
We then varied $\tau_1$ and $\tau_2$ to examine how the $\eta_{1500}$ evolution changes with different SFHs.

\begin{figure}
	\includegraphics[width=\columnwidth]{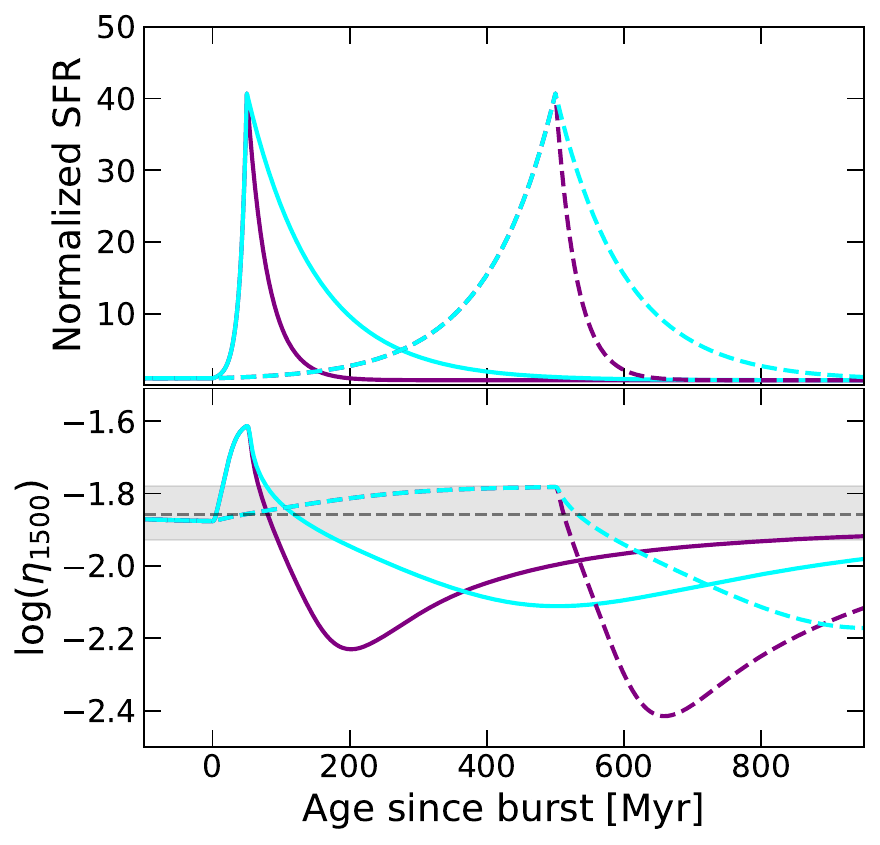}
    \caption{The $\eta_{1500}$ evolution under different SFH assumptions.
    The SFH is parameterized to model a rise and declining of SFR (Equation \ref{eqn:asymmetric_exp}), and we show four different set of rising/declining timescale (top panel).
    The resulting $\eta_{1500}$ evolutions are shown in the bottom panel.
    The shaded region and black dashed line present the range of $\eta_{1500}$ values expected in the steady constant SFHs as shown in Figure \ref{fig:eta_dist}.
}
    \label{fig:eta_FSPS_apdx}
\end{figure}

Figure \ref{fig:eta_FSPS_apdx} shows the $\eta_{1500}$ evolutions under four different SFHs, each of which represents fast rising and fast declining SF (solid purple: $(\tau_1,\tau_2)=(10,30)$ Myr), fast rising and slow declining (solid cyan: $(\tau_1,\tau_2)=(10,100)$ Myr), slow rising and fast declining (dashed purple: $(\tau_1,\tau_2)=(100,30)$ Myr), and slow rising and slow declining (dashed cyan: $(\tau_1,\tau_2)=(100,100)$ Myr). respectively.
If we fix the rising timescale and vary the declining timescale (i.e., comparing purple and cyan), the minimum $\eta_{1500}$ value changes.
On the other hand, if we vary the declining timescale (i.e., comparing solid and dashed lines), the maximum $\eta_{1500}$ value is mainly affected.
Therefore, we can roughly estimate the rising and declining timescale ($\tau_1,\tau_2$) by comparing the corresponding maximum/minimum $\eta_{1500}$ values with the observed $\eta_{1500}$ PDFs (Figure \ref{fig:eta_PDF}).

Both of interacting and non-interacting galaxy PDFs have a peak at high-$\eta_{1500}$ region, and a short rising SF timescale is required for the both.
On the other hand, the $\eta_{1500}$ PDF difference in the low-$\eta_{1500}$ region shown in Figure \ref{fig:eta_PDF} should point the difference in the declining timescale.
Namely, we found that the rising timescale $\tau_1$ for both galaxy sample should be shorter than $\lesssim100$ Myr to make a peak above the equilibrium range: the declining timescale $\tau_2$ for interacting galaxy sample should be shorter than $<100$ Myr as well to make another peak around $\log(\eta_{1500})<-2.1$, whereas that for non-interacting galaxies should be longer than $\gtrsim100$ Myr because of the lack of lowest-$\eta_{1500}$ population in the sample.
As a representative, in Figure \ref{fig:eta_FSPS}, we adopted $(\tau_1,\tau_2)=(10, 30)$ Myr model for interacting galaxies (red curves in the figure) and $(\tau_1,\tau_2)=(10,200)$ Myr model for non-interacting galaxies (blue curves).
It is worth noting that these values used for the toy models are only rough and qualitative estimations, and further detailed analysis is required to quantitatively determine these parameters as we discuss in Section \ref{sec:discussion}.

\subsection{Burst amplitude}
For simplicity, we here fixed the amplitude factor (e.g., parameter $A$), but the amplitude is known to affect the $\eta_{1500}$ distribution as well.
Thus one may worry that the $\eta_{1500}$ PDF difference between interacting/non-interacting galaxy sample can stem from the difference of the typical amplitude and not from the characteristic timescales.
In this subsection we give a further discussion on this point.

\citet{Emami2019} proposed the width of H$\alpha$ luminosity ($L_{\rm H\alpha}$) distribution around the main-sequence is useful as the proxy of the burst amplitude.
The more scattered $L_{\rm H\alpha}$ distribution indicates the higher amplitude.
Using both of $L_{\rm H\alpha}$ and $\eta_{1500}$ distribution is expected to break the degeneracy between the burst amplitude and the timescale.

\begin{figure}
	\includegraphics[width=\columnwidth]{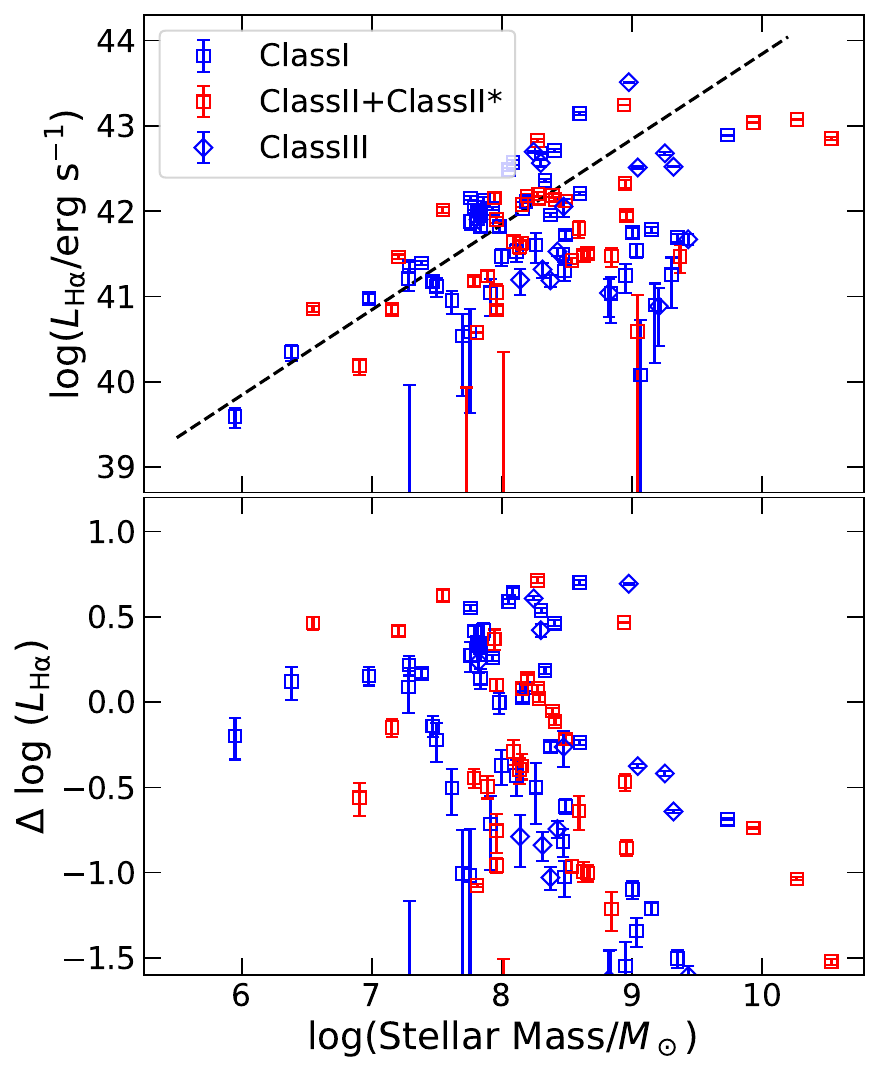}
    \caption{\textit{Top}: stellar mass vs. H$\alpha$ luminosity plot for the sample galaxies in this work.
    Symbols are the same as Figure \ref{fig:eta_dist}: red (blue) plots represent (non-)interacting galaxies.
    The black dashed line shows the mean trend of $L_{\rm H\alpha}$ against the stellar mass.
    \textit{Bottom}: stellar mass vs. $\Delta \log(L_{\rm H\alpha})$ plot. The offsets are measured from the mean trend shown in the top panel.
}
    \label{fig:Ha_luminosity}
\end{figure}

To this end, we measured the $\Delta \log(L_{\rm H\alpha})$, the offset from the (stellar mass dependent) average, for each sample galaxy in this work, and compare the standard deviation of $\Delta \log(L_{\rm H\alpha})$ in interacting and non-interacting galaxy sample.
We first calculated the H$\alpha$ luminosity for each sample galaxy from the photometric redshift $z_{\rm ml}$ and H$\alpha$ flux derived in Section \ref{subsec:Ha_meas}.
We then estimated the mean trend of $\log(L_{\rm H\alpha})$ against the stellar mass (black dashed line in Figure \ref{fig:Ha_luminosity} top), and subtract this trend from the H$\alpha$ luminosity of each galaxy to obtain the offset $\Delta \log(L_{\rm H\alpha})$ (Figure \ref{fig:Ha_luminosity} bottom).
We calculated the standard deviations of $\Delta \log(L_{\rm H\alpha})$ each in interacting and non-interacting galaxy sample, and obtained 0.415 and 0.457, respectively.
This result indicates that the width of the $L_{\rm H\alpha}$ distribution in interacting galaxy and non-interacting galaxy sample is similar, and the typical amplitude $A$ should not be different between the two samples.
Thus, the amplitude $A$ cannot explain the observed difference in the $\eta_{1500}$ PDF, and timescales of $\tau_1$ and $\tau_2$ should be the cause of difference.

\bsp	
\label{lastpage}
\end{document}